\documentclass[apj]{emulateapj}

\usepackage{amssymb, amsmath, apjfonts, natbib, color}

\def\degree{{\circ}}

\newdimen\digitwidth
\setbox1=\hbox{0}
\digitwidth=\wd1
\catcode`"=\active
\def"{\kern\digitwidth}

\slugcomment{ApJ accepted}

\begin{document}

\title{Vertical Phase Mixing across the Galactic Disk}

\author{Zhao-Yu Li\altaffilmark{1, 2}}

\altaffiltext{1}{Department of Astronomy, School of Physics and Astronomy, Shanghai Jiao Tong University, 800 Dongchuan Road, Shanghai 200240, China; correspondence should be addressed to: lizy.astro@sjtu.edu.cn}
\altaffiltext{2}{Key Laboratory for Particle Astrophysics and Cosmology (MOE) / Shanghai Key Laboratory for Particle Physics and Cosmology, Shanghai 200240, China}

\begin{abstract}

By combining the {\it LAMOST} and {\it Gaia} data, we investigate the vertical phase mixing across the Galactic disk. Our results confirm the existence of the phase space snail shells (or phase spirals) from 6 to 12 kpc. We find that grouping stars by the guiding radius ($R_{g}$), instead of the present radius ($R$) further enhances the snail shell signal in the following aspects: (1) clarity of the snail shell shape is increased; (2) more wraps of the snail shell can be seen; (3) the phase spaces are less affected by the lack of stars closer to the disk mid-plane due to extinction; (4) the phase space snail shell is amplified in greater radial ranges. Compared to the $R$-based snail shell, the quantitatively measured shapes are similar, except that the $R_{g}$-based snail shells show more wraps with better contrast. These lines of evidence lead to the conclusion that the guiding radius (angular momentum) is a fundamental parameter tracing the phase space snail shell across the Galactic disk. Results of our test particle simulations with impulse approximation verify that particles grouped according to $R_{g}$ reveal well-defined and sharper snail shell features. By comparing the radial profiles of the pitch angle between observation and simulation, the external perturbation can be constrained to $\sim$500$-$700 Myr ago. For future vertical phase mixing study, it is recommended to use the guiding radius with additional constraints on orbital hotness (ellipticity) to improve the clarity of the phase snail.

\end{abstract}

\keywords{Galaxy: disk --- Galaxy: kinematics and dynamics --- Galaxy: structure --- stars: kinematics and dynamics}

\section{INTRODUCTION}
\label{sec:intro}

The Milky Way, as a massive pure disk galaxy, is not in dynamical equilibrium. The current status of the Galactic disk is mainly shaped via both the internal and external mechanisms together. Resonances from the bar and spiral arms have been shown to induce observed fine structures in the velocity phase space \citep{dehnen_00, fux_01, antoja_etal_09, antoja_etal_11, quille_etal_11, hun_bov_18} and the large scale bulk motions in the Galactic disk \citep{sieber_etal_11, sieber_etal_12, carlin_etal_13, debatt_14, faure_etal_14, sun_etal_15, monari_etal_15, monari_etal_16, tian_etal_17, wang_etal_18a, wang_etal_18b}. The external perturbations from satellite galaxies or sub-halos can generate warps, flares, vertical density asymmetries or high order velocity modes in the Galactic disk, such as the bending and breathing vertical motions \citep{hun_too_69, quinn_etal_93, kazant_etal_08, purcel_etal_11, gomez_etal_13, willia_etal_13, donghi_etal_16, laport_etal_18a, laport_etal_18b}.

External perturbations also excite vertical oscillation and phase mixing of stars in the Galactic disk. According to \citet{widrow_etal_12}, the vertical density profile of the Galactic disk shows clear asymmetry with wave-like patterns (also see \citealt{ben_bov_19} for more recent update, and \citealt{an_19} from the chemical perspective). As first shown in \citet{antoja_etal_18} with {\it Gaia} DR2, stars near the solar radius exhibit a snail shell feature in the $Z-V_{Z}$ phase space, representing the ongoing vertical phase mixing that probably happened $300 - 900$ Myr ago. This is a 2D representation of the signal discovered in \citet{widrow_etal_12}. The snail shell forms as a result of anharmonic oscillation of stars in the vertical direction; the vertical oscillation frequency ($\Omega_{Z}$) is anti-correlated with the vertical action ($J_{Z}$) and the angular momentum ($L_{Z}$), i.e., $V_{\phi}$ for stars in the solar neighborhood \citep{bin_sch_18}. The phase space snail shell can be recognized for stars in a wide range of ages \citep{tian_etal_18, laport_etal_19}, and different chemical properties \citep{blandh_etal_19}. By dissecting stars in the $V_{R}-V_{\phi}$ phase space into distinct arches, \citet{li_she_20} found clear snail shells only in the stars on dynamically colder orbits, i.e., stars with $|V_{\phi} - V_{\rm LSR}| \leq 30$ km/s (or $J_{R} < 0.04$ kpc$^2$/Myr). The hotter orbits, on the other hand, may have phase wrapped away already due to the much larger radial excursions to facilitate faster phase mixing. The absence of the clear snail shells on hotter orbits suggests that the vertical perturbation occurred at least 500 Myr ago. However, different opinions also exist for the origin of the vertical phase mixing process \citep[see][]{khoper_etal_19a, ben_bov_20}.

The phase space snail shell has also been found beyond the solar neighborhood in the Milky Way disk. Based on the {\it Gaia} DR2 data around the solar neighborhood (within 1 kpc), slightly different phase space snail shells are found at different radius, suggesting the snail shell as a signature of corrugated waves propagating through the disk \citep{blandh_etal_19}. \citet{laport_etal_19} later explored larger radial ranges from 6.6 to 10 kpc, and confirmed the existence of the snail shell at these locations. Similar results can also be seen in \citet{wang_etal_19} based on the {\it LAMOST} and {\it Gaia} DR2 data from 6.34 to 12.34 kpc. These studies arrive at similar results; the snail shell at larger radius is more elongated (squashed) along the $Z$ ($V_{Z}$) direction and less wound, due to the weaker vertical restoring force and longer orbital time period at larger radius. Recently, \citet{xu_etal_20} further link the $Z-V_{Z}$ phase space distributions at different radius with the azimuthal bulk motions across the Galactic disk out to 15 kpc, indicating tight connection between the external perturbation and the disk kinematics.

For a volume limited sample, the angular momentum ($L_{Z}$), i.e., the guiding radius ($R_{g}$), may help to better reveal the global kinematic patterns. For example, based on a local sample near the solar neighborhood, the Galactic warp was identified with the positive correlation with the wave-like pattern between the mean vertical velocity and the guiding radius \citep{sch_deh_18, huang_etal_18}. Recently, \citet{fri_sch_19} found that the dependence of the mean radial motion on the angular momentum (guiding radius) also shows wave-like behavior. The shape of the $V_{R} - R_{g}$ profile is similar to, but much stronger than the $V_{R}-R$ profile; patterns in $V_{R}-R$ profile have probably been washed out by excursions of stars around their guiding centers \citep{fri_sch_19}. \citet{khanna_etal_19} found that the snail shell shape remains the same at different radius (within 1 kpc from the Sun) for stars with the same $L_{Z}$, but varies at different $L_{Z}$. They suggested that the angular momentum could be a more robust quantity in characterizing the snail shell.

Instead of mapping the stellar kinematics and tracing the phase space maps in the traditional spatial volume, grouping stars according to the guiding radius may be another promising way to visualize the morphological or kinematic structures in the disk \citep[e.g.,][]{khoper_etal_20} (but see \citealt{hunt_etal_20} for a different point of view). Aiming to probe the phase space snail shell in a large radial range in the Galactic disk with good number statistics, we combine the {\it Gaia} radial velocity sample (RVS) and the {\it LAMOST} DR6 sample together with all the stars in both samples included, i.e., {\it LAMOST} DR6 $\cup$ {\it Gaia} RVS. Although the {\it LAMOST} survey does not sample the sky as uniformly as the {\it Gaia} survey, it has good spatial coverage in the Galactic Anti-Center direction with radial velocity information for fainter stars at larger distances, compensating the relatively brighter {\it Gaia} RVS sample. With the combined large sample, we can investigate the spatial variation of the snail shell with angular momentum in detail.\footnote{\citet{wang_etal_19} has shown that the $Z-V_{Z}$ phase space distributions are consistent between the {\it Gaia} RVS and {\it LAMOST}-{\it Gaia} cross-matched samples from 6 to 12 kpc, which is also verified by our tests. The combination of the two samples should help to enhance the phase space signal.}

\section{SAMPLE}
\label{sec:samp}

\begin{figure}
\epsscale{1.15}
\plotone{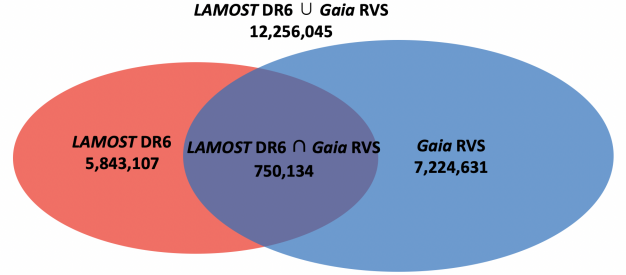}
\caption{Number of stars in the {\it LAMOST} DR6 and {\it Gaia} RVS samples used in this work. The combined main sample ({\it LAMOST} DR6 $\cup$ {\it Gaia} RVS) contains 12,256,045 stars in total. The cross-matched sample ({\it LAMOST} DR6 $\cap$ {\it Gaia} RVS) contains 750,134 stars (used in Fig.~\ref{fig:rv_cmp}). For the main sample ({\it LAMOST} DR6 $\cup$ {\it Gaia} RVS), after crossing-match with the Bayesian distance catalog from \citet{bailer_etal_18} and removing stars with large velocity uncertainties, the final sample contains 11,350,423 stars.}
\epsscale{1.0}
\label{fig:sample}
\end{figure}

{\it LAMOST} DR6 provides reliable radial velocity values for 5,843,107 stars, focusing on the Galactic Anti-Center direction \citep{zhao_etal_12, deng_etal_12, liu_etal_14}. The {\it Gaia} RVS sample contains 7,225,631 stars with accurate position and velocity information \citep{gaia_etal_18a}. To achieve good statistics, we utilize all the stars to form a main sample of 12,256,045 stars ({\it LAMOST} DR6 $\cup$ {\it Gaia} RVS). Fig.~\ref{fig:sample} summarizes the number of stars in the two samples.

\begin{figure}
\epsscale{1.2}
\plotone{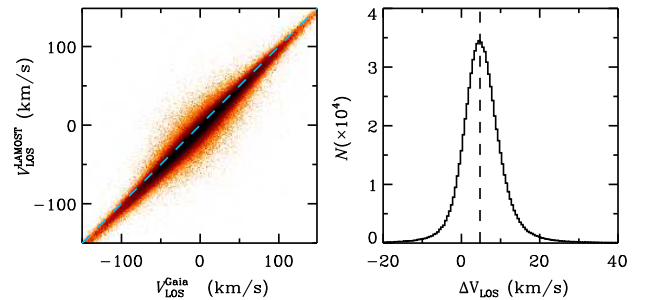}
\caption{Comparison between the observed line-of-sight velocities for the {\it LAMOST} DR6 and {\it Gaia} RVS common stars, with the correlation between $V_{\rm LOS}^{\rm LAMOST}$ and $V_{\rm LOC}^{\rm Gaia}$ in the left panel, and the distribution of their velocity difference ($\Delta V_{\rm LOS} = V_{\rm LOS}^{\rm Gaia} - V_{\rm LOS}^{\rm LAMOST}$) in the right panel. The cyan dashed line in the left panel represents the one-to-one relationship. The {\it LAMOST} radial velocity is systematically lower than the {\it Gaia} radial velocity by 4.75 km/s (vertical dashed line in the right panel).}
\epsscale{1.0}
\label{fig:rv_cmp}
\end{figure}

There are 750,134 stars in common between {\it LAMOST} DR6 and {\it Gaia} RVS samples ({\it LAMOST} DR6 $\cap$ {\it Gaia} RVS). Fig.~\ref{fig:rv_cmp} compares the line-of-sight velocities observed in {\it Gaia} ($V_{\rm LOS}^{Gaia}$) and {\it LAMOST} ($V_{\rm LOS}^{LAMOST}$) for these common stars. As shown in the left panel, the agreement between the velocities are quite good. However, there exists a small systematic offset, with the $V_{\rm LOS}^{LAMOST}$ slightly smaller than $V_{\rm LOS}^{Gaia}$. The distribution of the velocity difference ($\Delta V_{\rm LOS} = V_{\rm LOS}^{Gaia} - V_{\rm LOS}^{LAMOST}$) is shown in the right panel. The peak position of the distribution is marked with the vertical dashed line. The line-of-sight velocity observed in {\it LAMOST} is systematically lower than the {\it Gaia} radial velocity by 4.75 km/s.\footnote{This systematic velocity offset in {\it LAMOST} data has been noticed in previous works \citep{tian_etal_15, sch_aum_17}.} In the following analysis, for these common stars, only the {\it Gaia} measured velocities are adopted. For the rest of the stars in the {\it LAMOST} DR6 sample, we have added 4.75 km/s to the line-of-sight velocity to compensate this effect.

\begin{figure}
\epsscale{1.2}
\plotone{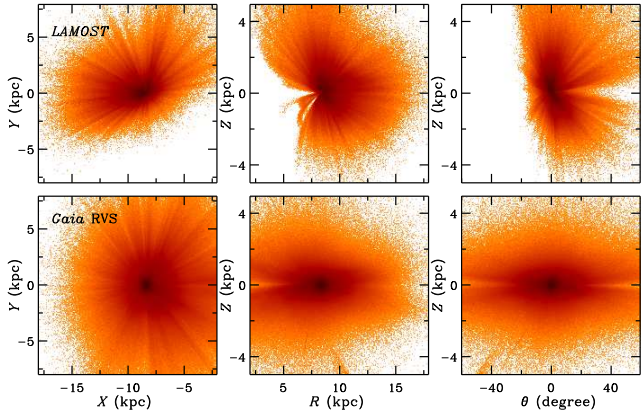}
\caption{Spatial distributions of the {\it LAMOST} DR6 (top) and {\it Gaia} RVS samples (bottom) in the $X-Y$ plane (left column), the $R-Z$ plane (middle column), and the $\theta-Z$ plane (right column). Compared to the relatively brighter {\it Gaia} RVS sample, {\it LAMOST} sample has its advantage of more faint stars at relatively larger distances.}
\epsscale{1.0}
\label{fig:sample_space}
\end{figure}

We adopt the Bayesian distance from \citet{bailer_etal_18} for the whole sample. There are 12,094,719 stars in the main sample with the Bayesian distance determined. Similar to previous works, we choose $(X_{\odot},\ Y_{\odot},\ Z_{\odot})$ = ($-$8.34, 0, 0.027) kpc as the Sun position \citep{reid_etal_14}. The local standard of rest (LSR) circular velocity $V_{\rm LSR}$ is set to 240 km/s \citep{reid_etal_14}. Here the peculiar velocities of the Sun with respect to LSR are set to $(U_{\odot}, V_{\odot}, W_{\odot})$ = (11.1, 12.24, 7.25) km/s \citep{schonr_12}.\footnote{The main results are not affected if we choose other values of the solar peculiar motion, e.g., \citet{tian_etal_15} or \citet{huang_etal_15}.}

The spatial distributions of the {\it LAMOST} and {\it Gaia} RVS samples are shown in Fig.~\ref{fig:sample_space}. The {\it Gaia} RVS sample distributions are quite symmetric in the $Y$, $Z$ and $\theta$ directions.\footnote{$\theta$ is the angle in the Galactocentric polar coordinate which increases clockwise ($\theta=0^\degree$ for the Sun$-$Galactic center line).} On the other hand, the {\it LAMOST} survey mainly covers the Galactic Anti-Center direction and $\theta > 0$ in the positive $Y$ axis. Since the limiting magnitude is deeper in the {\it LAMOST} spectroscopic survey, it also extends further in the vertical direction than the {\it Gaia} survey.

\begin{figure}
\epsscale{1.3}
\plotone{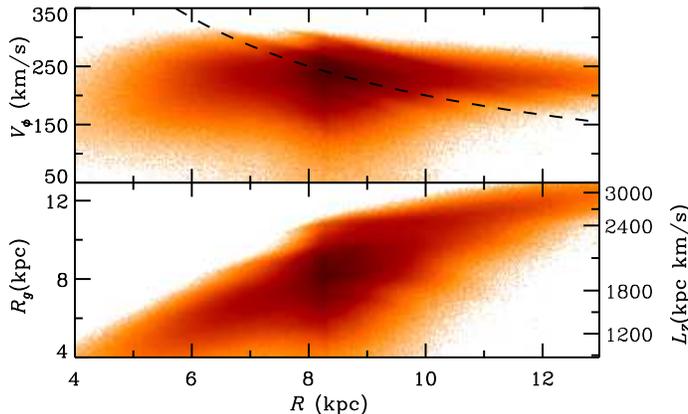}
\caption{Number density distributions of our main sample in $R-V_{\phi}$ (top) and $R-R_{g}(L_{Z})$ phase spaces (bottom). The top panel shows several prominent diagonal ridges consistent with previous works. The dashed curve represents the constant angular momentum at $L_{Z} = 2000 $ kpc km/s. In the bottom panel, there are several nearly horizontal stripes, suggesting roughly constant angular momentum of stars at different radial ranges.}
\epsscale{1.0}
\label{fig:rvp}
\end{figure}

We further remove stars with relative distance uncertainties larger than 25\%, velocity uncertainties larger than 50 km/s, and absolute velocities ($|V_{R}|, |V_{\phi} - V_{\rm LSR}|$, and $|V_{Z}|$) larger than 400 km/s. The final sample contains 11,350,423 stars in total. In the following analysis, we will focus on the radial range from 6 to 12 kpc for both the Galactocentric radius ($R$) and the guiding radius ($R_{g}$). In order to enhance the visualization of the phase space structures, at 6 and 7 kpc, the radial annulus width is set to 0.6 kpc (i.e., $\pm 0.3$ kpc), while the annulus width is 0.4 kpc (i.e., $\pm 0.2$ kpc) at the other radii. The typical velocity uncertainty is about 1 km/s for the radial, azimuthal, and vertical velocities \citep{gaia_etal_18a, antoja_etal_18}. 

The guiding radius ($R_{g}$) of each star is calculated according to the rotation curve in \citet{huang_etal_16}. As shown in Fig.~12 in \citet{huang_etal_16}, the rotation curve is roughly flat at $\sim$ 240 km/s (within $\sim$ 25 kpc) with prominent wiggles at 11 and 20 kpc. Adopting the values of the rotation curve in Table~3 of \citet{huang_etal_16}, the angular momentum corresponding to the circular motion at each radius (i.e., the guiding radius) can be derived to get the $L_{Z}-R_{g}$ profile. Then the guiding radius of each star in our sample is found by mapping the observed angular momentum to the derived $L_{Z}-R_{g}$ profile.

To highlight the snail shell feature, we adopt the number density contrast map used in \citet{laport_etal_19} and \citet{li_she_20}. To evaluate the influence of the parallax bias in the {\it Gaia} catalog, we also test the same analysis with the parallax corrected {\it Gaia} sample \citep{schonr_etal_19}, which contains 6,565,715 stars with relatively smaller spatial coverage. The main results in this work are robust and not affected.

\section{Phase Spaces across the Galactic Disk}

In this section, we explore several different phase spaces of the sample, namely, the $R-V_{\phi}$, $R-R_{g}$, $V_{R} - V_{\phi}$, and the $Z-V_{Z}$ phase spaces. As we will show later, major structures in these phase spaces are consistent with previous studies, confirming the robustness of the sample to trace the kinematic status of the Milky Way disk.

\subsection{$R-V_{\phi}$ and $R-R_{g}$ Phase Spaces}

Fig.~\ref{fig:rvp} shows the number density maps of the main sample in the $R-V_{\phi}$ (top) and $R-R_{g} (L_{Z})$ phase spaces (bottom). The top panel reveals several diagonal ridge lines, consistent with the previous observations \citep{antoja_etal_18, laport_etal_19, fragko_etal_19}. In the bottom panel, several nearly horizontal stripes can be seen across the phase space, suggesting a possible connection between the ridges in the $R-V_{\phi}$ phase space and the constant angular momentum curves. However, as shown in \citet{khanna_etal_19} and \citet{fragko_etal_19}, at higher $V_{\phi}$ the stripes may be more consistent with the constant energy lines. The observed diagonal ridges could have quite complicated origins, considering the presence of the ridge line structure in the $V_{R}$ or metallicity color-coded $R-V_{\phi}$ phase space \citep{fragko_etal_19, laport_etal_19, liang_etal_19, wang_etal_20}.

From the bottom panel, stars at certain guiding radius are actually located within a large radial range. At the radius far from the solar neighborhood, the number of stars can be improved by selecting stars according to their guiding radii with the inclusion of stars close to the Sun. This could help to enhance the phase space snail shell feature at a larger radial range.

\subsection{$V_{R} - V_{\phi}$ Phase Space}

\begin{figure*}
\epsscale{1.2}
\plotone{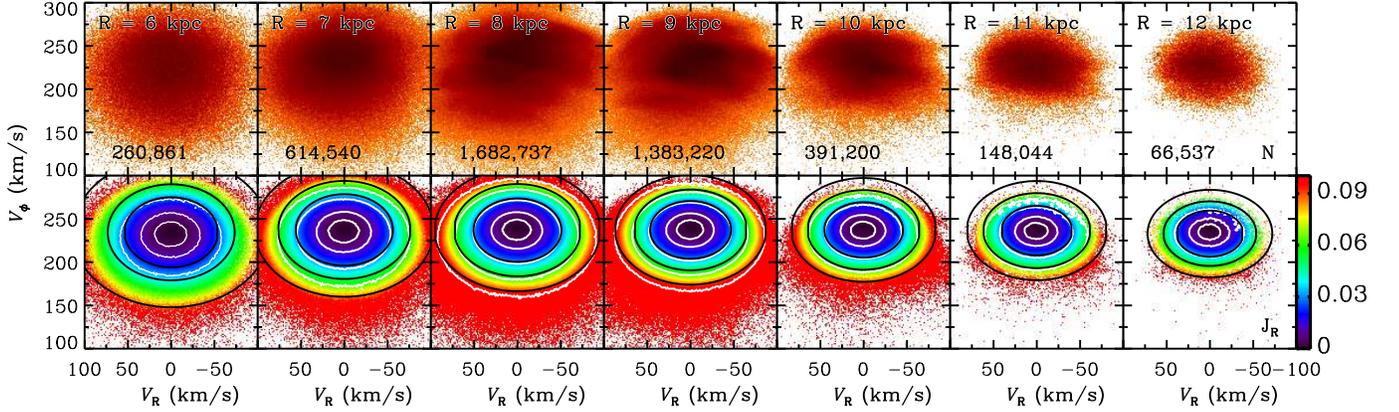}
\caption{$V_{R} - V_{\phi}$ phase space distributions of stars in different Galactocentric radial annulus from $R = 6$ to $12$ kpc. The top and bottom rows show the number density and the $J_{R}$ color-coded maps, respectively. In the top row, arches can be seen in different radius, especially at $R = 8, 9$, and 10 kpc. The number of stars in each radial range is also shown in the panel. In the bottom row, the white curves represent the constant $J_{R}$ contours, while the black ellipses are the analytical estimation of the constant $J_{R}$ contours based on the epicycle theory.}
\epsscale{1.0}
\label{fig:uv_r}
\end{figure*}

\begin{figure*}
\epsscale{1.2}
\plotone{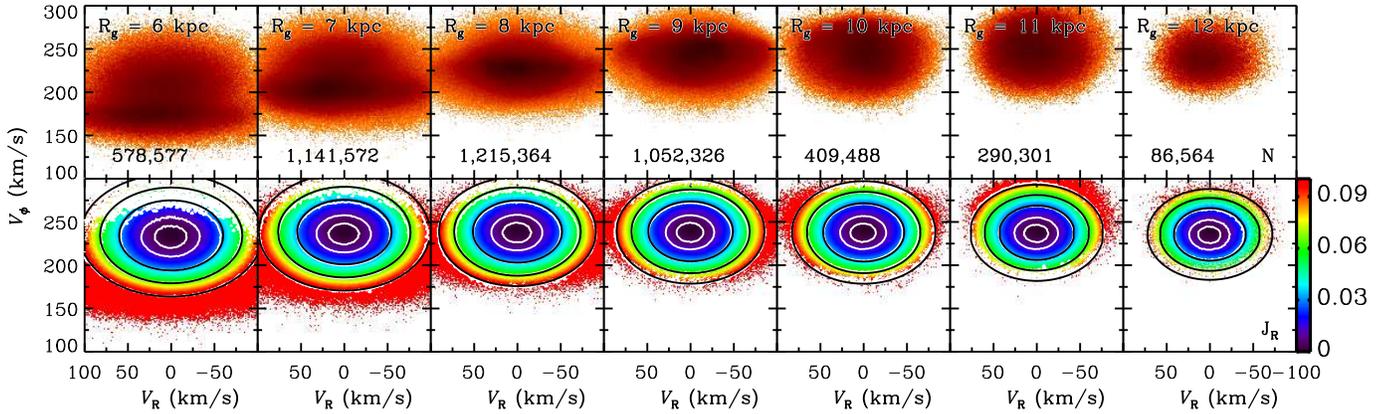}
\caption{$V_{R} - V_{\phi}$ phase space distributions of stars in different guiding radius $R_{g}$ annulus from $6$ to $12$ kpc. The layout of this figure is the same with Fig.~\ref{fig:uv_r}. The distributions are relatively smooth without clear arches. The white contour curves and black ellipses have slightly larger ellipticity than those in Fig.~\ref{fig:uv_r}.}
\epsscale{1.0}
\label{fig:uv_rg}
\end{figure*}

Fig.~\ref{fig:uv_r} shows the $V_{R}-V_{\phi}$ phase space of stars at different radius, with the number density and the $J_{R}$ color-coded maps in the top and bottom rows, respectively. From the main sample, we study the velocity phase space distributions at the 7 radial annuli centered at $R = 6, 7, 8, 9, 10, 11,$ and 12 kpc with the same width as mentioned before. {\it Gaia} DR2 has revealed the new arch-like structures in the $V_{R}-V_{\phi}$ phase space for stars in the solar neighborhood, enclosing the classical moving groups \citep{gaia_etal_18b, antoja_etal_18, li_she_20}. In fact,  similar arch-like features can be seen at $R = 8, 9,$ and 10 kpc. As the radius increases, the arches seem to systematically shift towards lower $V_{\phi}$, consistent with previous observational studies \citep{antoja_etal_14, ramos_etal_18}. 

According to the epicycle theory, at a given radius, the azimuthal velocity difference between a star and the local circular velocity can be estimated as $V_{\phi}(R_0) - V_{\rm circ}(R_0) = \kappa x / \gamma$, where $\kappa$ is the epicycle (radial oscillation) frequency, $x$ is the radial displacement from the guiding radius, and $\gamma = 2\Omega_{g} / \kappa$. Under the flat rotation curve assumption, $\gamma$ equals to $\sqrt{2}$. The ratio between the maximum radial velocity and the maximum azimuthal velocity difference can be estimated, which is simply $1/\gamma$. Adopting the best-fit potential of Model I from \citet{irrgan_etal_13}, we use AGAMA package \citep{vasili_19} to derive the epicycle frequencies and the action values. In Fig.~\ref{fig:uv_r}, from left to right panels (6 to 12 kpc), the $\gamma$ values are 1.32, 1.35, 1.39, 1.40, 1.41, 1.43, and 1.44, respectively. In the bottom panels for the $V_{R} - V_{\phi}$ phase spaces color-coded with $J_{R}$, the black ellipses with this axis ratio are overplotted compared to the white curves of the constant $J_{R}$ contours. The white contours and the black ellipses agree well with each other.

Similarly, we choose 7 different guiding radial annuli from 6 to 12 kpc. The $V_{R}-V_{\phi}$ phase space in different $R_{g}$ ranges are shown in Fig.~\ref{fig:uv_rg} with the same layout as Fig.~\ref{fig:uv_r}. Compared to Fig.~\ref{fig:uv_r}, the distribution is relatively smooth without clear arches\footnote{An elongated overdensity can be recognized at $R_{g} = 6$ kpc with $V_{\phi} \approx 170$ km/s, and at $R_{g} = 7$ kpc with $V_{\phi} \approx 200$ km/s. These stars mainly come from the solar radius.}. Similar to Fig.~\ref{fig:uv_r}, the azimuthal velocity difference between a star and the circular velocity at its guiding radius can be estimated as $V_{\phi}(R_0) - V_{\rm circ}(R_{g}) = \gamma \kappa x/2$.\footnote{$V_{\phi}(R_0) - V_{\rm circ}(R_{g})=\Omega_{0}R_{0}-\Omega_{g}R_{g}\simeq\Omega_{g}(R_{0}-R_{g})=\Omega_{g}x=\gamma\kappa x/2.$} Therefore, the ratio between the maximum radial and azimuthal velocity difference is simply $\gamma/2$. The $\gamma$ values from $R_{g} = 6$ to 12 kpc are 1.35, 1.37, 1.39, 1.41, 1.43, 1.45, and 1.46, respectively. The axis ratio is slightly smaller than that in Fig.~\ref{fig:uv_r}. The black ellipses with this axis ratio are shown in the bottom panels, with good agreement with the white contours.

\subsection{$Z-V_{Z}$ Phase Space}

\begin{figure*}
\epsscale{1.}
\plottwo{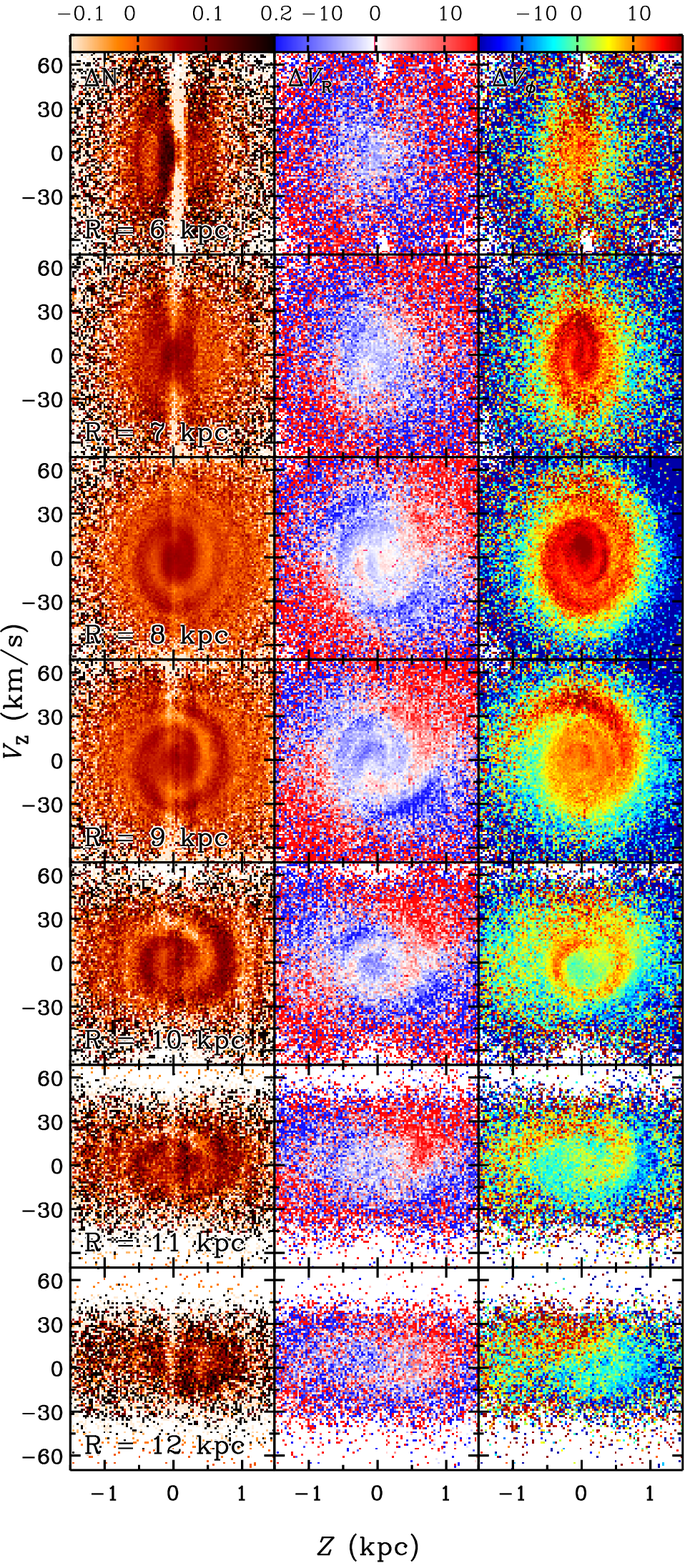}{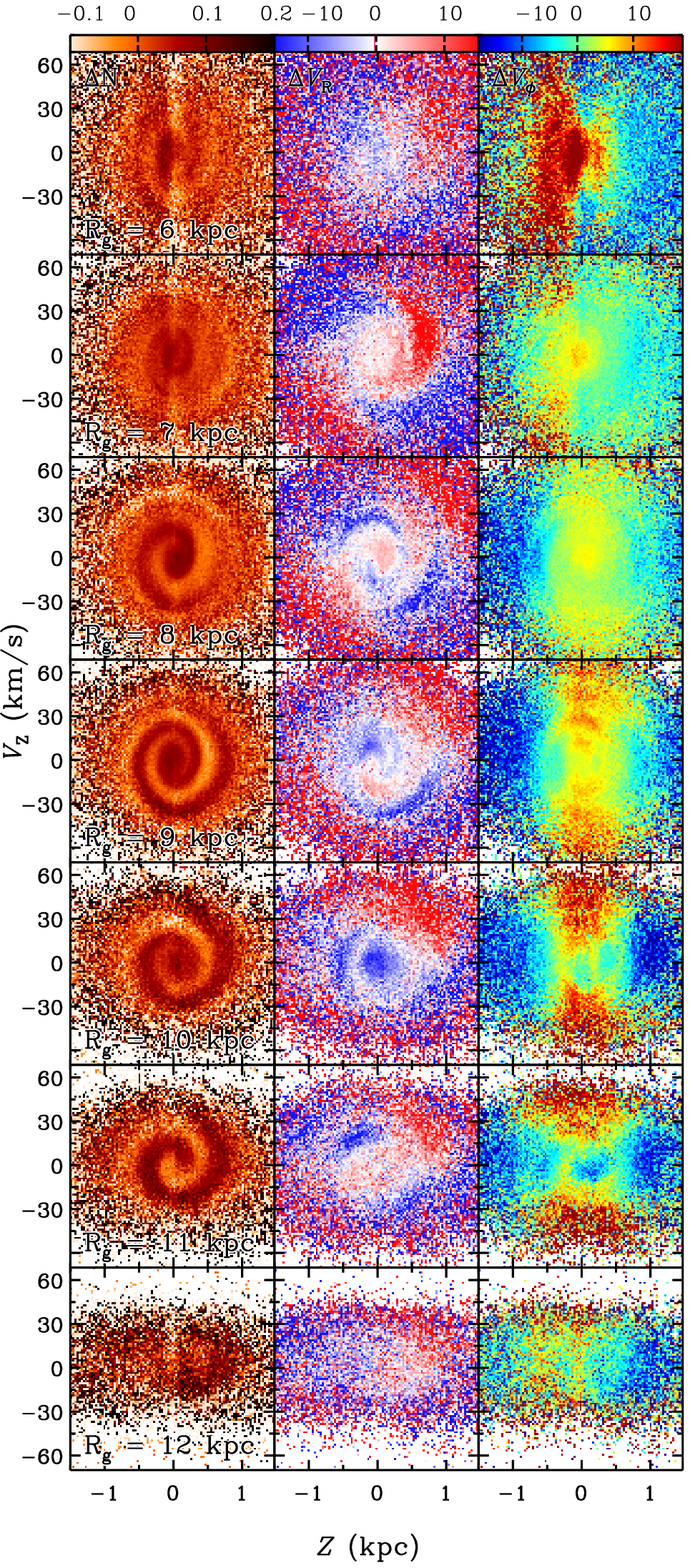}
\caption{The $Z-V_{Z}$ phase space distributions of the main sample from 6 kpc (top) to 12 kpc (bottom) in different Galactocentric radial annulus (left figure) and guiding radial annulus (right figure). In the left (right) figure, the first, second, and third columns show the number density contrast $\Delta N$ map, $V_{R}$, and $V_{\phi}$ color-coded phase spaces, respectively. Each pixel corresponds to $0.03\ {\rm kpc} \times 1.4\ {\rm km/s}$. At each radius, the snail shells are more clearly visualized in the $\Delta N$ map in the guiding radial annulus in the right figure. Vertical stripes due to the observational bias become much less noticeable in the $R_{g}$-based snail shell.}
\epsscale{1.0}
\label{fig:snail}
\end{figure*}

In Fig.~\ref{fig:snail}, the $Z-V_{Z}$ phase spaces of stars at different Galactocentric radii ($R$) are shown in the left figure, with the first, second and third columns representing the number density contrast ($\Delta N$), $V_{R}$ and $V_{\phi}$ color-coded maps, respectively. The number density contrast is derived as $\Delta N = N/\widetilde{N} - 1$, where $\widetilde{N}$ is the Gaussian kernel convolved number density distribution\footnote{Before the convolution, we have taken the logarithm of the original number density map to enhance the fine structures in the phase space.} \citep{laport_etal_19, li_she_20}. The snail shell can be recognized in all the panels, confirming the snail shell feature as a global phenomenon across the Galactic disk. The shapes of the snail shells as revealed by the $\Delta N$ map vary at different radius. As the radius increases, the snail shell becomes more elongated along the $Z$ axis, more squashed along the $V_{Z}$ axis, and less wound, due to the decreasing vertical restoring force and oscillation frequency at larger radius. These results are consistent with \citet{laport_etal_19} and \citet{wang_etal_19}. 

The snail shells at 6, 7, and 12 kpc are relatively weak and hard to discern in $\Delta N$ maps, although at $R = 7$ kpc, the $V_{\phi}$ color-coded phase space still reveals a snail shell pattern. At $R = 8$ kpc, the snail shell in $\Delta N$ map looks roughly similar to those in the $V_{\phi}$ color-coded map, which is not true for $R = 9$ and 10 kpc. As already demonstrated by \citet{li_she_20}, the snail shell appeared in the $V_{\phi}$ color-coded phase space may not truly reflect the real phase mixing signal due to the variation of the snail shell shape at different $V_{\phi}$ \citep[also see][]{laport_etal_19}. 

The $R_{g}$-based $Z-V_{Z}$ phase space maps are shown in the right figure of Fig.~\ref{fig:snail}. Clear snail shells can be seen in the $\Delta N$ maps, which look similar to the snail shells in the corresponding $R$-based phase spaces with improved clarity. On the other hand, there is no clear snail shell-like pattern in the $V_{R}$ or $V_{\phi}$ color-coded phase space shown in the second and third columns. Patterns in the $V_{\phi}$ color-coded phase spaces are discussed more in Appendix A. In Appendix B, we show $R_{g}$-based phase spaces for 30 sequential annuli, which are evenly sampled between 6 and 12 kpc with 0.2 kpc width. A gradual variation of the snail shell shape towards larger radius is quite clear. In addition, in the $R_{g}$-based phase space, the gap in the $R$-based phase space around $Z=0$ (lack of stars due to dust extinction) is also mitigated. In fact, the angular momentum space is not complete in terms of star counts. It is biased towards stars in the solar neighborhood where the stellar number density is the highest; at lower (higher) $R_{g}$, more stars in the lower (higher) $V_{\phi}$ tail in the velocity distributions are captured in the $R_{g}$ selected subsamples. Nonetheless, they still show coherent phase mixing signals to enhance the snail shell feature at each $R_{g}$. More details are shown in Appendix C. 

\begin{figure*}
\epsscale{1.}
\plotone{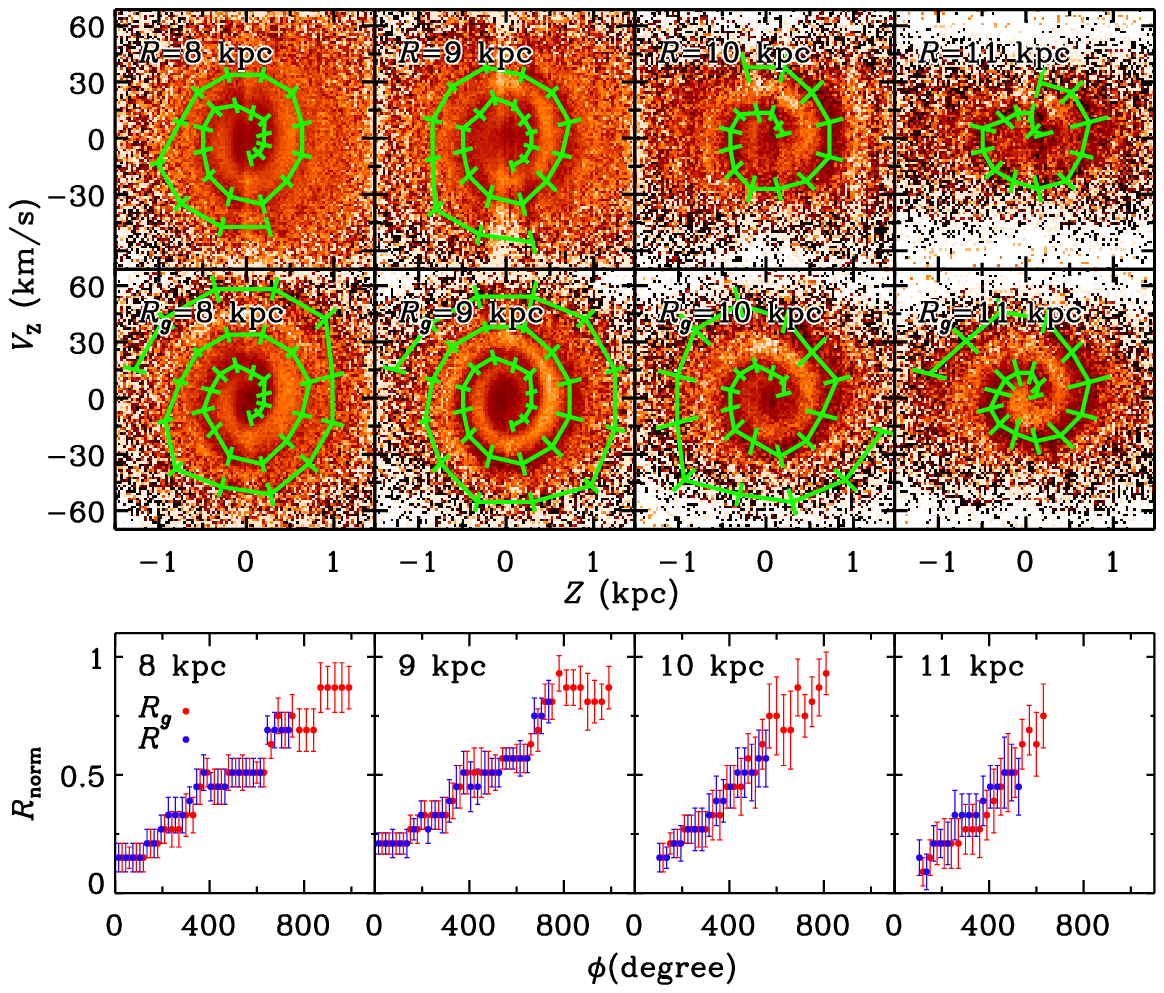}
\caption{Phase space snail shells shapes measured at different $R$ (top row) and $R_{g}$ (middle row) from 8 to 11 kpc. The overlaid green curves with error bars represent the extracted profiles of the snail shells. Apparently, compared to the top row, more wraps of the shell can be identified in the middle row at each $R_{g}$. The bottom row compares the phase space snail shell shapes between $R$ (blue) and $R_{g}$ (red)  from 8 to 11 kpc. $\phi$ and $R_{\rm norm}$ are the angle (increasing counterclockwise in the phase space) and normalized dimensionless radius of the measured snail shell profiles in the polar coordinates. The snail shell shapes at the same $R$ and $R_{g}$ are consistent, except that the profiles in $R_{g}$ extend further with larger $\phi$ which indicates more wraps of the snail shell in the phase space.}
\epsscale{1.0}
\label{fig:snail_rrg}
\end{figure*}

\begin{figure}
\epsscale{1.1}
\plotone{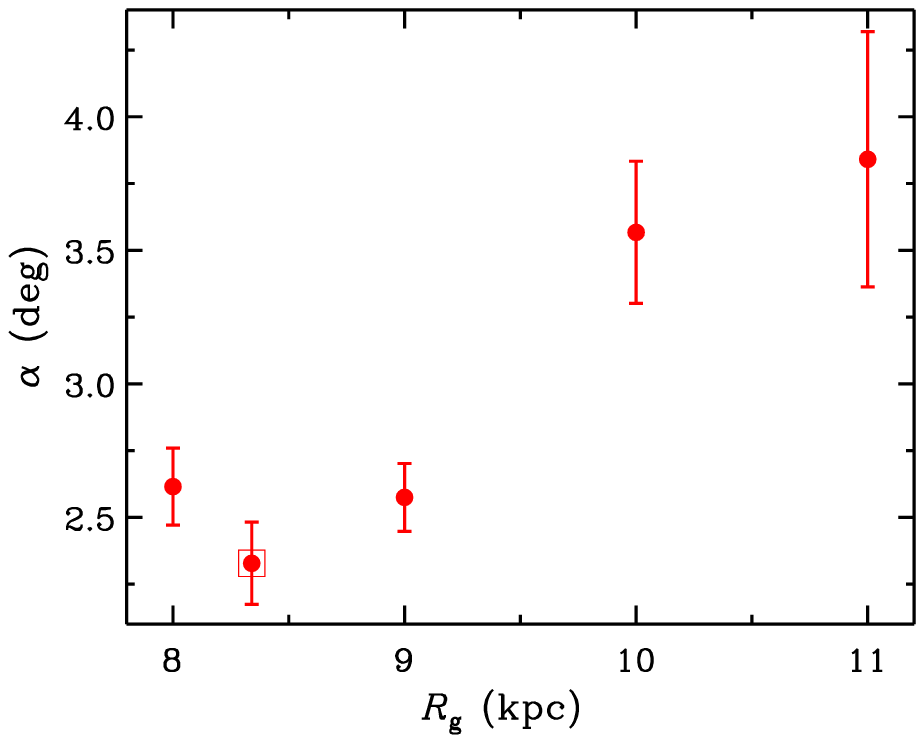}
\caption{The pitch angle $\alpha$ at different guiding radii $R_{g}$. The snail shells at the outer disk show larger pitch angle (i.e., loosely wound) compared to the snail shell in the inner disk. This is consistent with the theoretical expectation. The red point with a square denotes the value of the solar neighborhood.}
\epsscale{1.}
\label{fig:pitch_ang_prof}
\end{figure}

\section{Angular Momentum: A Fundamental Parameter to Trace the Snail Shell}

In the epicycle approximation, a star revolves around the guiding center which is on a circular orbit around the Galactic center. Grouping stars according to the guiding radius instead of the present Galactocentric radius can avoid the  mixture of stars with different angular momentum. In this section, we perform quantitative comparisons of the phase space snail shells between different $R$ and $R_{g}$ ranges, and compare to other previous works to emphasize the importance of guiding radius to trace the vertical phase mixing signal.

\subsection{Snail Shell Shape Measurement}

As shown in the $\Delta N$ maps of the $R$- and $R_{g}$-based phase spaces in Fig.~\ref{fig:snail}, the snail shell is reflected with dark-brown color (local maximum) separated by light-brown color (local minimum). Ideally, if we work in the polar coordinate of the phase space, and to choose a wedge centered at (0, 0), then averaging the signal at each phase space radius\footnote{In the phase space, $Z$ and $V_{Z}$ values at each point are divided by 1.5 kpc and 70 km/s, respectively, to get a normalized dimensionless radius $R_{\rm norm} = \sqrt{(Z/1.5 {\rm kpc})^2 + (V_{Z}/70 {\rm km/s})^2}$.} in the wedge will result in a $\Delta N$ profile, with the peaks and troughs corresponding to the snail shell and inter shell regions, respectively. After identifying the shell positions in each wedge, the identified peak positions in all the wedges can be simply connected to recover the whole snail shell structure. The phase space is equally divided into 12 wedges with $30^\degree$ azimuthal angle. The phase space angle $\phi = 0^\degree$ is defined in the direction of $Z = 0$ and $V_{Z} < 0$ that increases counterclockwise following the snail shell outwards from the central region of the phase space.

\begin{figure*}
\epsscale{1.}
\plotone{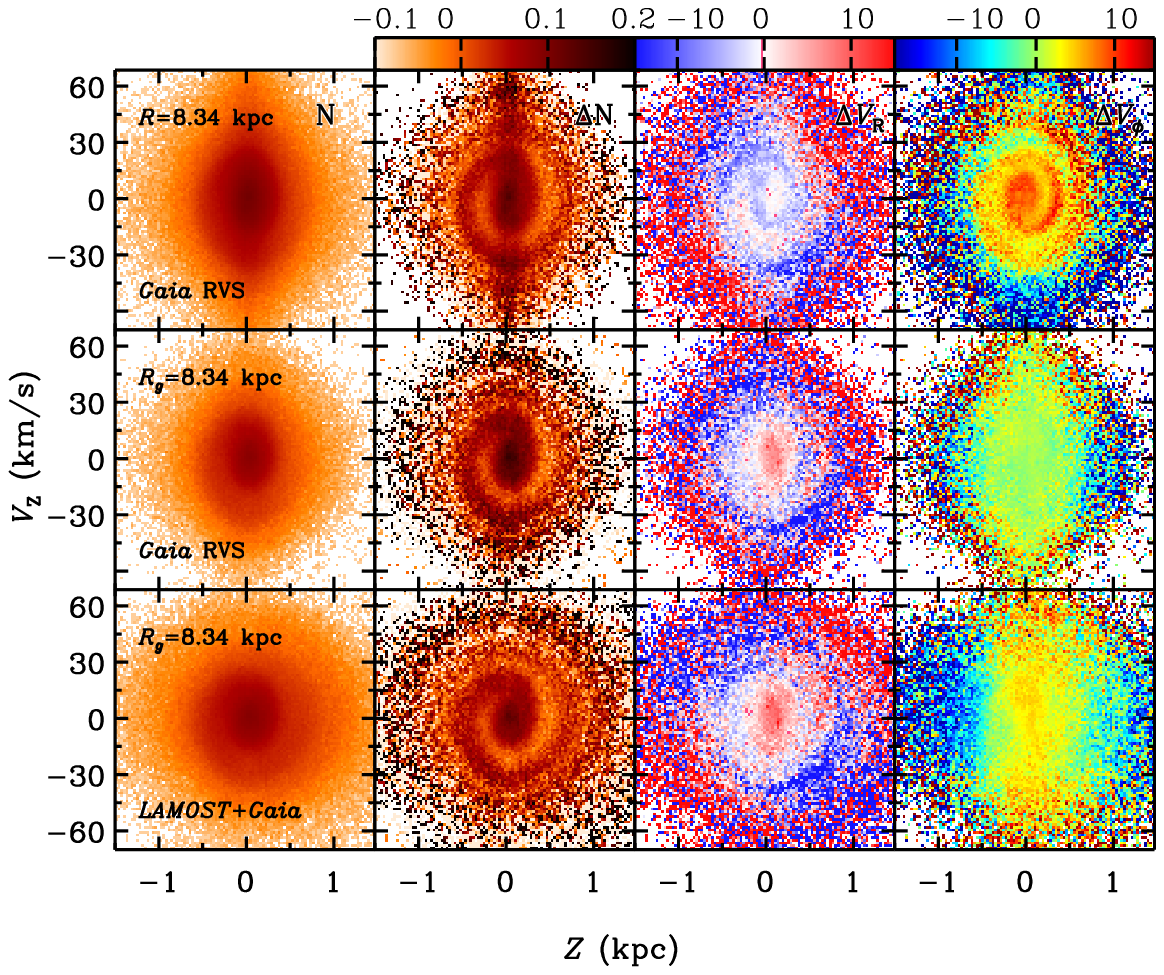}
\caption{The $R$- and $R_{g}$-based phase space maps at the solar radius ($R, R_{g} = 8.34$ kpc) using the {\it Gaia} RVS and the main sample (similar to Fig.~\ref{fig:snail}). The $R$- and $R_{g}$-based phase space snail shells look similar with improved clarity and one more wrap seen in the middle and bottom rows in $R_{g}$-based phase space.}
\epsscale{1.0}
\label{fig:snail_rrg_SN}
\end{figure*}

We have identified the snail shell structure in the $\Delta N$ maps at $R, R_{g} = 8, 9, 10$, and 11 kpc, which are shown in Fig.~\ref{fig:snail_rrg} with the measured snail shell shape overlaid on the $\Delta N$ map.\footnote{The snail shells at 6, 7, and 12 kpc are relatively weak and hard to measure quantitatively.} The outermost wrap of the snail shell is truncated depending on its relative amplitude. At 8 and 9 kpc, the snail shell is measured up to $\phi = 720^\degree$, then the measurement is terminated when the snail shell relative amplitude\footnote{The relative amplitude of the snail shell is estimated by averaging the $\Delta N$ differences between the local peak position (the snail shell) and the adjacent local troughs (the inter shell regions on the two sides of each snail shell).} is below 0.025. For 10 and 11 kpc, the snail shell is traced to $\phi = 540^\degree$. Then the profile is truncated when the snail shell relative amplitude is lower than 0.1. The error bar of the snail shell shape is determined by 1/4 separation between the two troughs on each side of the local peak (except for the outer most shell, which is determined by 1/2 separation between the peak and the inner trough since no outer trough can be determined). As shown in Fig.~\ref{fig:snail_rrg}, the green curves with error bars generally follow the visually identified snail shell. Due to the large fluctuations in the signal, the outer shell is relatively difficult to map compared to the inner shell. Clearly, the green curve marking the $R_{g}$-based snail shell shows more wraps and well defined shape than the $R$-based snail shell. 

The bottom row of Fig.~\ref{fig:snail_rrg} shows the shape of the $R$- and $R_{g}$-based snail shell in the $R_{\rm norm} - \phi$ plane. At each radius, the two types of points follow the same track, suggesting that the snail shell shapes are similar. The red dot ($R_{g}$) extends to larger $\phi$ and $R_{\rm norm}$ values to show more wraps in the phase space. In addition, the slope in $R_{\rm norm} - \phi$ plane is steeper at larger radius compared to smaller ones. This is consistent with the theoretical expectation that snail shell at larger radius is more loosely wound than the snail shell at the smaller radius.

To quantify the degree of woundedness of the phase space snail shell, the concept of pitch angle $\alpha$ is adopted\footnote{The pitch angle $\alpha$ can be calculated via $\displaystyle \cot{(\alpha)} = \left| R \frac{d \phi}{d R} \right|$.}, which is often used in spiral arm measurement. By fitting a linear function between $R_{\rm norm}$ and $\phi$ in the bottom row of Fig.~\ref{fig:snail_rrg}, the pitch angle at each guiding radius can be estimated. The result is shown in Fig.~\ref{fig:pitch_ang_prof}. As expected, the pitch angle increases from $\sim 2.5^\degree$ to $4^\degree$ with $R_{g}$ increasing from 8 to 11 kpc.

\subsection{Comparison with Previous Works}

We first compare our results with \citet{antoja_etal_18}, the discovery paper of the phase space snail shell. As shown in the top row of Fig.~\ref{fig:snail_rrg_SN}, we reproduce the phase space snail shell with stars near the solar radius in the {\it Gaia} RVS sample ($R = 8.34 \pm 0.1$ kpc), similar to Fig.~1 in \citet{antoja_etal_18}. For comparison, in the middle row, we also generate the $R_{g}$-based snail shell using {\it Gaia} RVS data. The bottom row shows the $R_{g}$-based snail shell using the {\it LAMOST + Gaia} sample. As shown in the $\Delta N$ maps (second column), it seems that the snail shell features are similar except that $R_{g}$-based snail shells are clear with one more wrap.

\begin{figure*}
\epsscale{1.}
\plotone{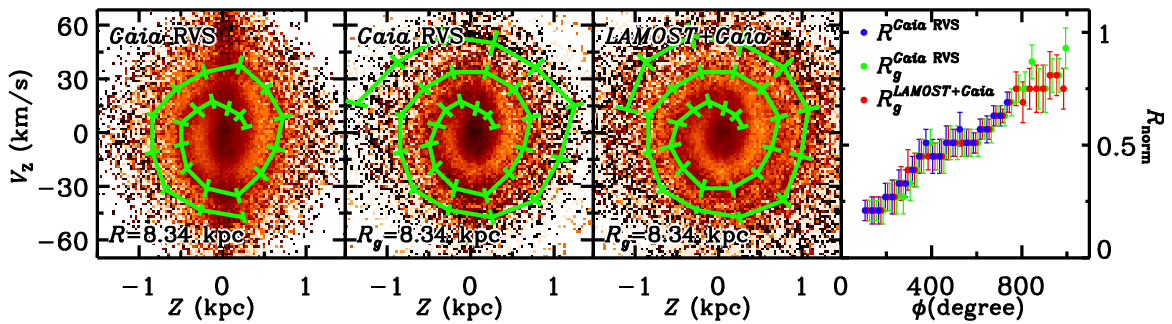}
\caption{Measured $R$- and $R_{g}$-based phase space snail shells in the solar neighborhood ($R, R_{g} = 8.34$ kpc) similar to Fig.~\ref{fig:snail_rrg}. The right panel compares the snail shapes shown in the left three panels, which follow the similar trend, with the $R_{g}$-based snail shell extending further with one more wrap.}
\epsscale{1.0}
\label{fig:sn_r_rg_shape}
\end{figure*}

\begin{figure*}
\epsscale{0.9}
\plotone{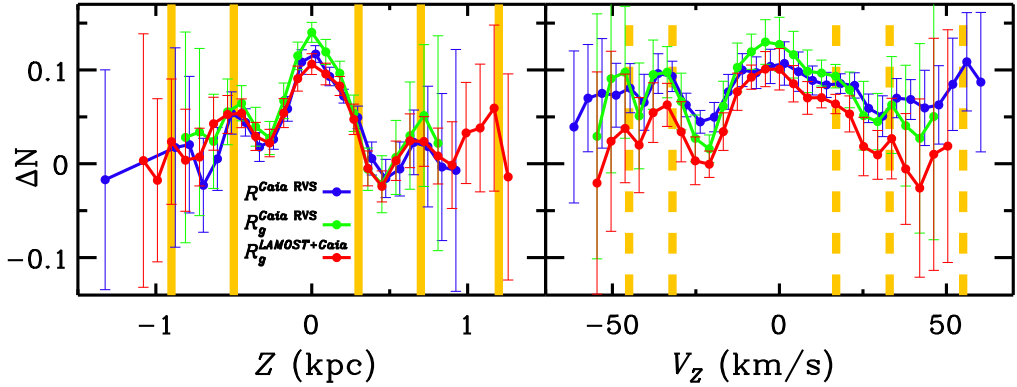}
\caption{Comparison of the $R$- and $R_{g}$-based phase space snail shell amplitudes in the solar neighborhood. The vertical yellow solid and dashed lines correspond to the turning-around ($V_{Z} = 0$) and mid-plane points ($Z = 0$) of the snail shell, respectively. The yellow lines are all located in the local maximum positions of the red and green lines in $R_{g}$-based profiles, with larger differences between the peak and the adjacent local minimum than the $R$-based profile (blue lines), indicating larger contrast/amplitude of the $R_{g}$-based snail shell (red and green lines).}
\epsscale{1.0}
\label{fig:sn_r_rg_amp}
\end{figure*}

\begin{figure*}
\epsscale{1.2}
\plotone{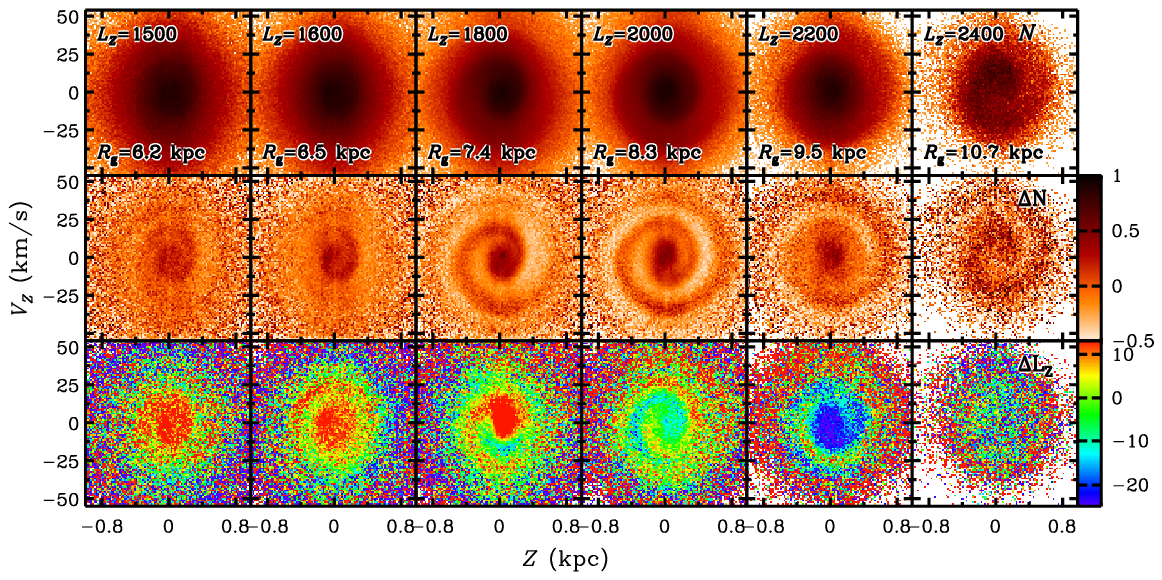}
\caption{The $Z-V_{Z}$ phase space distributions of stars in different angular momentum ranges. The median $L_{Z}$ increases from 1500 to 2400 kpc km/s from left to right columns, with the number density, density contrast, and $L_{Z}$ color-coded phase spaces from top to bottom rows. In each panel, the bin width $\Delta L_{Z}$ is 200 kpc km/s ($\pm 0.4$ kpc in $R_{g}$) to be consistent with \citet{khanna_etal_19}. Panels in the bottom row well reproduce Fig.~16 in \citet{khanna_etal_19}. The difference between the $L_{Z}$ color-coded phase space and the $\Delta N$ map is quite clear.}
\epsscale{1.0}
\label{fig:snail_lz_200}
\end{figure*}

The snail shell shapes of the two phase spaces are measured with the method mentioned in the previous section. The $\Delta N$ maps with the snail shell curves overlaid are shown in the left three panels in Fig.~\ref{fig:sn_r_rg_shape}. The comparison between the three curves are shown in the right panel. Clearly, the snail shell shapes are consistent within $\phi < 700^\degree$ (in the inner region of the phase space). The red and green dots ($R_{g}$) extend to larger $R_{\rm norm}$ and $\phi$ values.

\begin{figure*}
\epsscale{1.0}
\plottwo{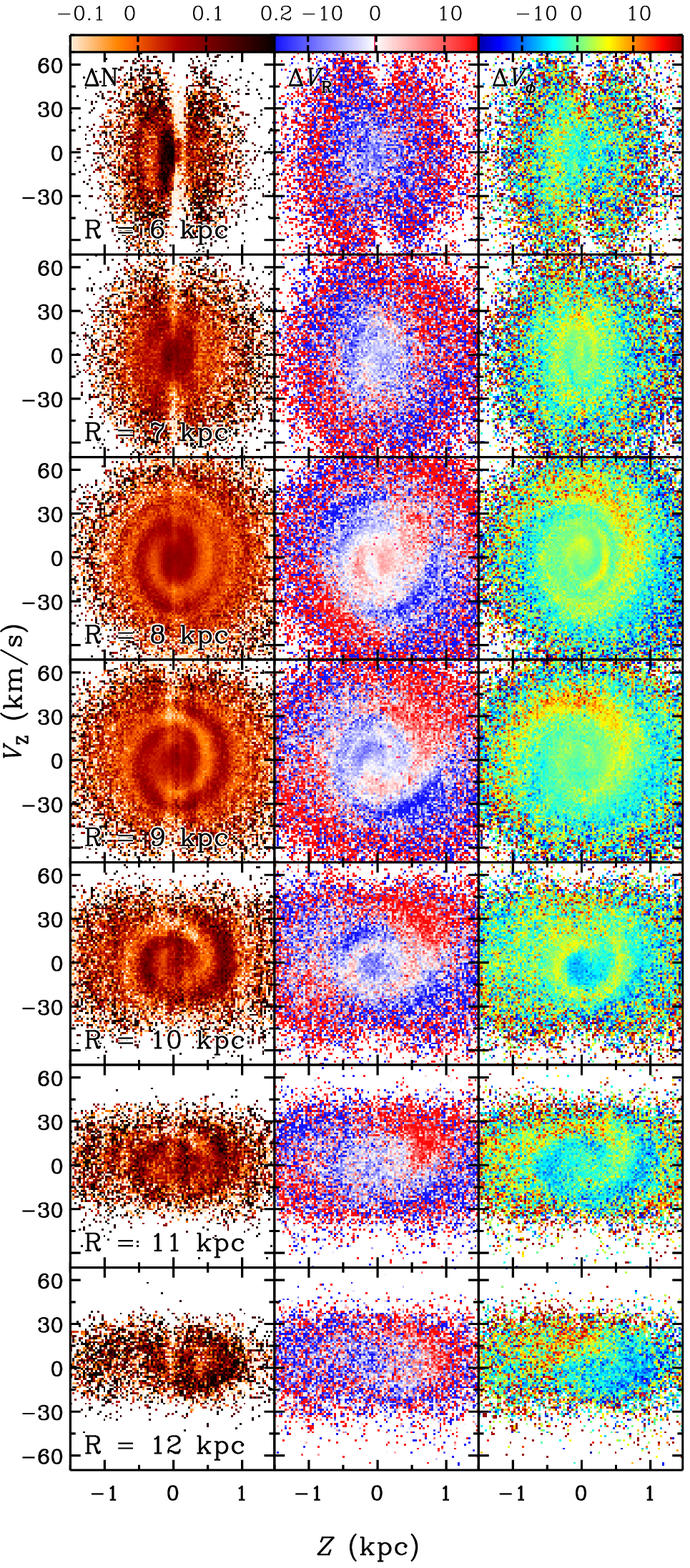}{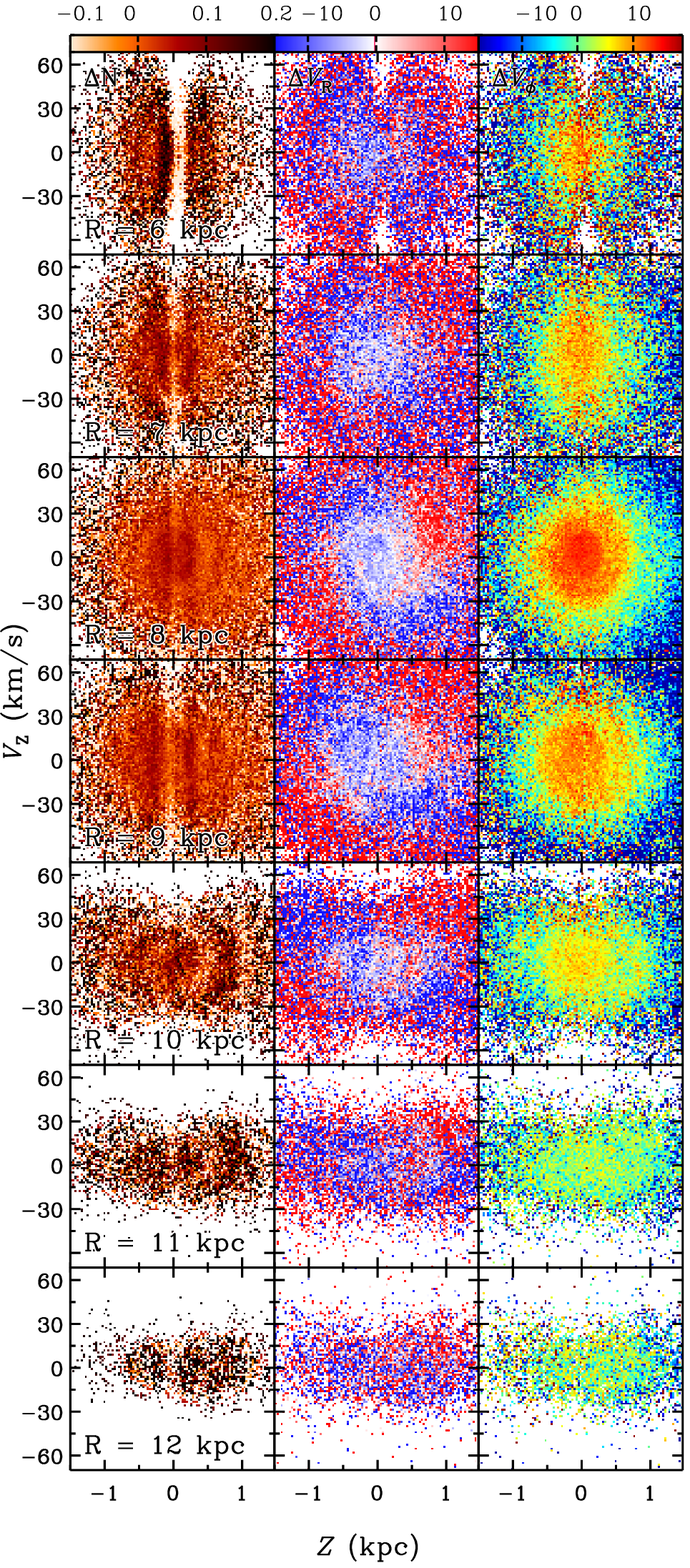}
\caption{The $Z-V_{Z}$ phase space distributions for the colder orbits (left figure) and hotter orbits (right figure) at different radial ranges. The layout of the figure is the same as Fig.~\ref{fig:snail}.}
\epsscale{1.0}
\label{fig:snail_cold_hot_r}
\end{figure*}

\begin{figure*}
\epsscale{1.}
\plottwo{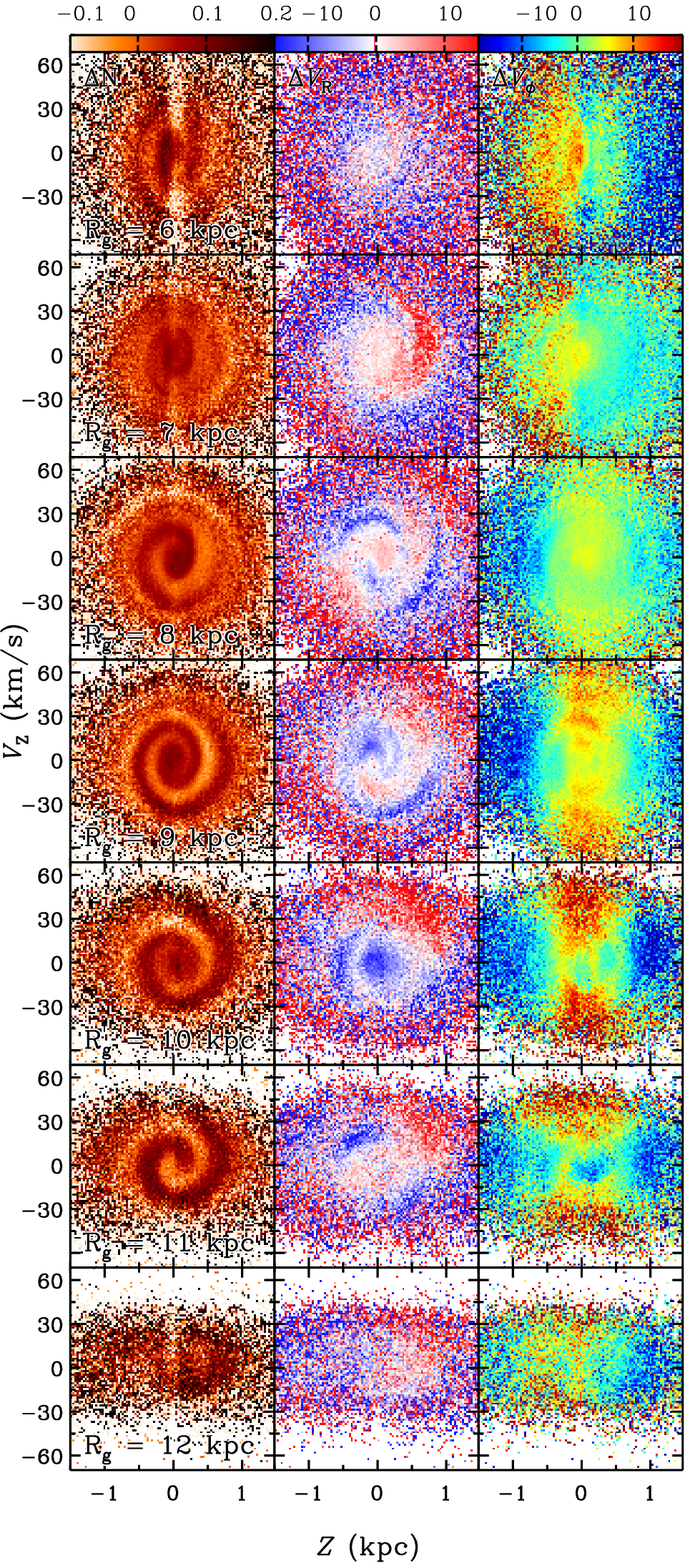}{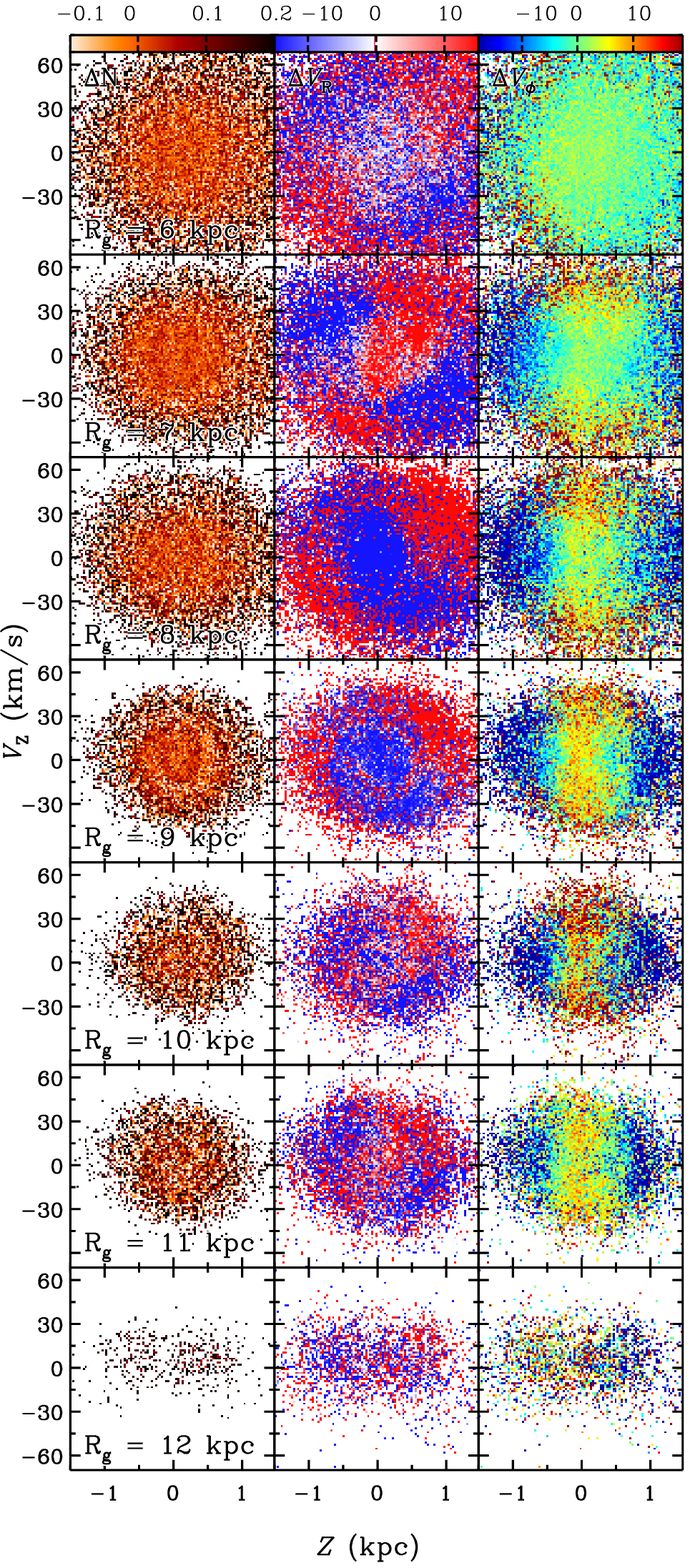}
\caption{The $Z-V_{Z}$ phase space distributions for the colder orbits (left figure) and hotter orbits (right figure) at different guiding radial ranges. The layout is the same with Fig.~\ref{fig:snail}. Note here we have raised the criteria of hotter orbits to $J_{R} > 0.06$.}
\epsscale{1.0}
\label{fig:snail_cold_hot_rg}
\end{figure*}

\begin{figure*}
\epsscale{1.}
\plotone{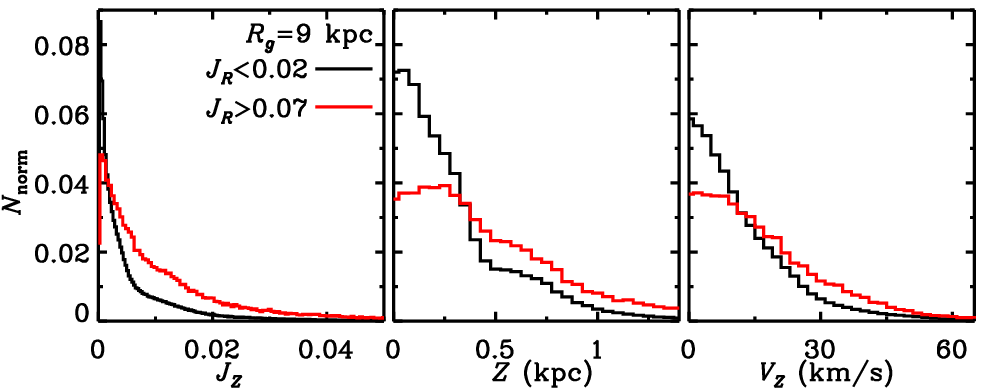}
\caption{Normalized distributions of $J_{Z}$, $Z$ and $V_{Z}$ for stars at $R_{g} = 9$ kpc. The black and red lines represent the stars with $J_{R} < 0.02$ and $J_{R} > 0.07$, respectively. The hotter stars with larger $J_{R}$ show larger dispersion in the vertical direction compared to the colder ones with smaller $J_{R}$.}
\label{fig:cold_hot_9}
\end{figure*}

The snail shell clarity is also estimated with the $\Delta N$ profiles extracted along $Z$ and $V_{Z}$ stripes; in the $\Delta N$ map, we choose the stripe with $|V_{Z}| < 15$ km/s to get the $\Delta N - Z$ profile, and the stripe with $|Z| < 0.15$ kpc to get the $\Delta N - V_{Z}$ profile. The results are shown in Fig.~\ref{fig:sn_r_rg_amp} with the left and right panels corresponding to the $\Delta N$ profiles along $Z$ and $V_{Z}$, respectively\footnote{The profiles are truncated with $\Delta N$ uncertainty larger than 0.14.}. The yellow vertical solid and dashed lines mark the snail shell turning-around points ($V_{Z} = 0$) and mid-plane points ($Z = 0$), respectively. The local peak position in the red and green curves ($R_{g}$) agree quite well with the yellow lines, with relatively larger differences between the peak and trough positions ($\sim$ 0.08). The blue curve ($R$) also seems to agree with the position of the snail shells, but with relatively smaller difference between the peaks and troughs ($\sim$ 0.05).  The comparison confirms that in the solar neighborhood, the snail shell in the $R_{g}$-based phase space is better revealed than the traditional $R$-based phase space.

\begin{figure*}
\epsscale{1.}
\plotone{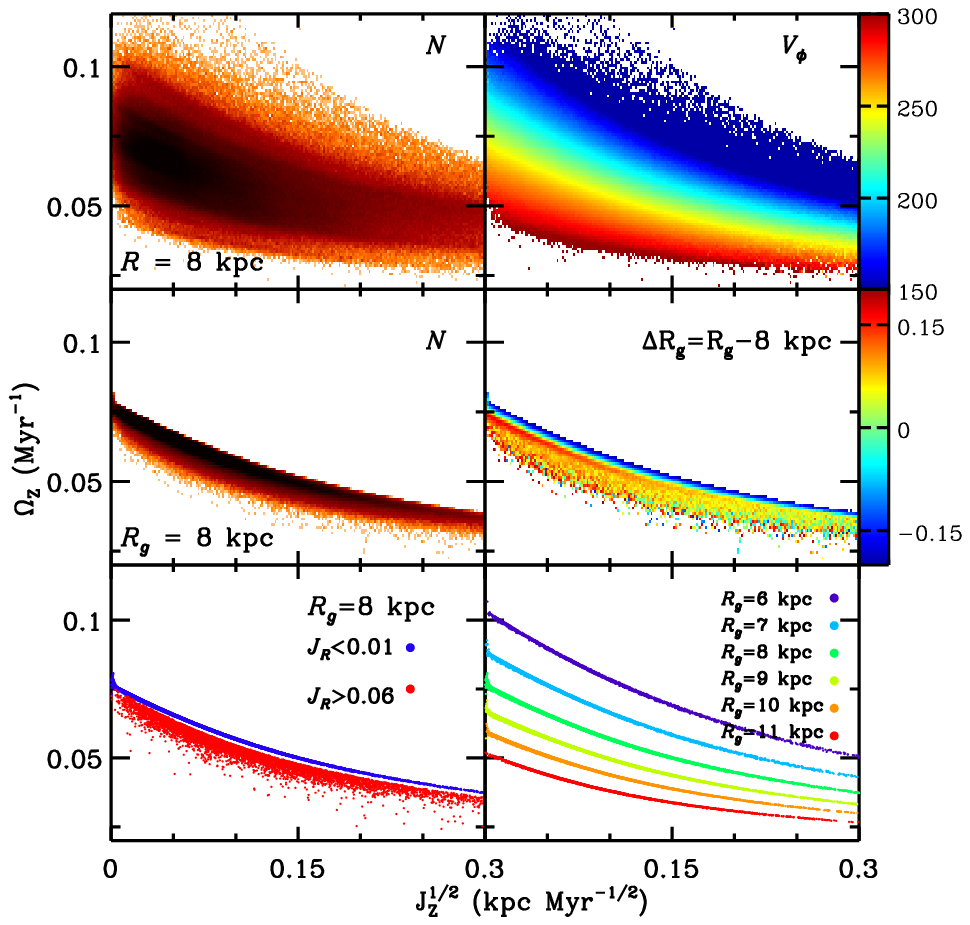}
\caption{$\Omega_{Z} - \sqrt{J_{Z}}$ distributions of stars at $R = 8$ kpc (top row) and $R_{g } = 8$ kpc (middle row). The top left and right panels show the number density and $V_{\phi}$ color-coded $\Omega_{Z} - \sqrt{J_{Z}}$ distributions at $R = 8$ kpc, respectively. The middle left and right panels show the number density and $\Delta R_{g} = R_{g} - 8$ kpc color-coded $\Omega_{Z} - \sqrt{J_{Z}}$ distributions, respectively. The distribution is much narrower for stars in different guiding radius bin (middle row), resulting in the clear phase space snail shell. At certain radius $R$, $\Omega_{Z}$ generally decreases with increasing $V_{\phi}$, producing the snail shell in the $V_{\phi}$ color-coded phase space. At $R_{g} = 8$ kpc, $\Omega_{Z}$ decreases slightly with increasing $\Delta R_{g}$. The bottom left panel shows the $\Omega_{Z} - \sqrt{J_{Z}}$ distribution of stars at $R_{g} = 8 \pm 0.01$ kpc with $J_{R} < 0.01$ (blue) and $J_{R} > 0.06$ (red). The bottom right panel compares the $\Omega_{Z} - \sqrt{J_{Z}}$ distributions of stars at different $R_{g}$ (from 6 to 11 kpc with $0.02$ kpc width annulus) with $J_{R} < 0.01$. Stars that closely follow circular motions at each guiding radius (with small $J_{R}$) show very tight correlations between $\Omega_{Z}$ and $\sqrt{J_{Z}}$.}
\epsscale{1.0}
\label{fig:omg_jz_n}
\end{figure*}

Based on a small sample covering $\sim$ 1 kpc distance from the Sun, \citet{khanna_etal_19} found that the phase spiral pattern for a given $L_{Z}$ is almost invariant with radius, suggesting the angular momentum as a more robust quantity to characterize the snail shell compared to $V_{\phi}$. As shown in Fig.~16 of \citet{khanna_etal_19}, the snail shell pattern revealed in the $L_{Z}$ color-coded phase space ($\Delta L_{Z} = 200$ kpc km/s) with the orientation of the snail shell changing with $L_{Z}$. Our result improves upon \citet{khanna_etal_19} in the aspect that we have explored a large radial range and performed comparisons of the $\Delta N$ phase spaces (representing the intrinsic shape of the snail shell rather than $V_{\phi}$ or $L_{Z}$ color-coded phase space) of different $R$ and $R_{g}$ ranges. According to our results, $R_{g}$ (or $L_{Z}$) is not only robust, but also fundamental to trace the phase space snail shell across the Galactic disk.

\begin{figure*}
\epsscale{1}
\plottwo{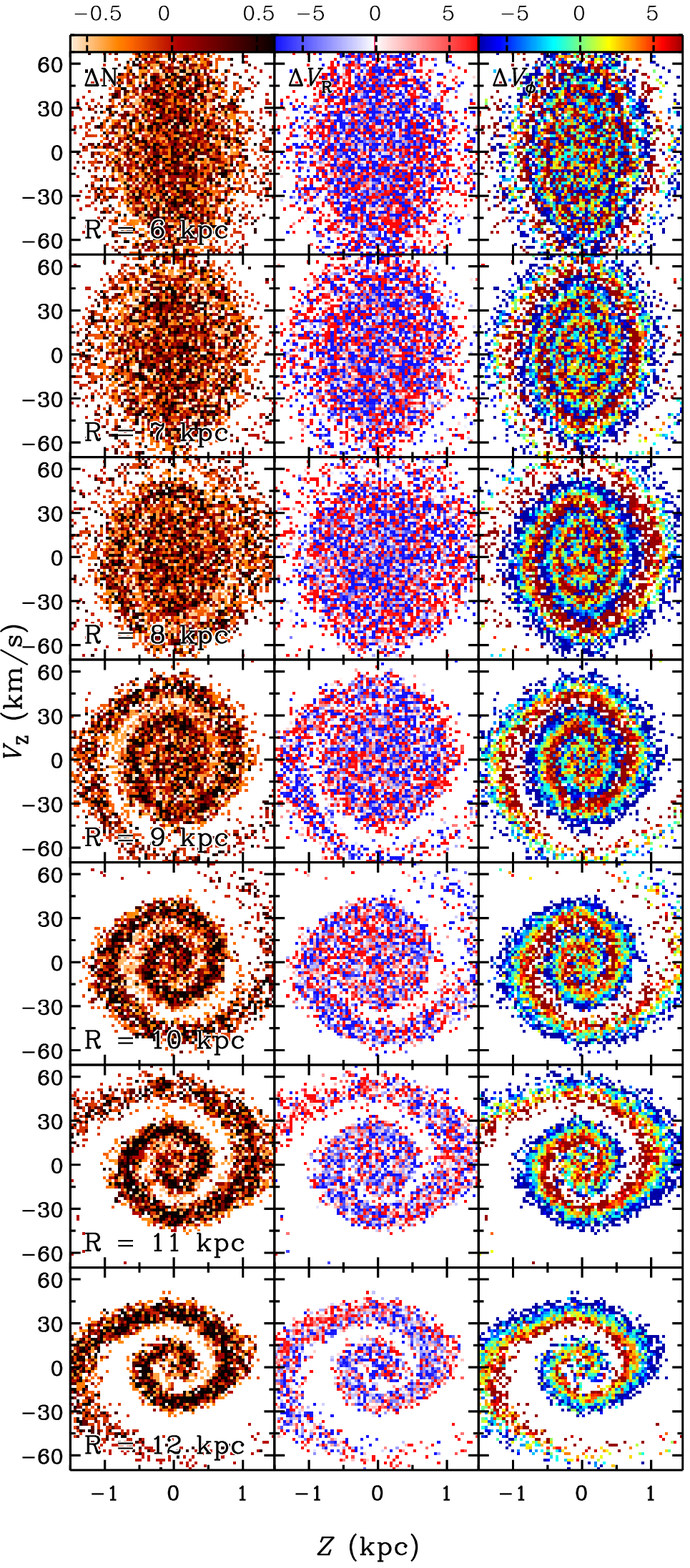}{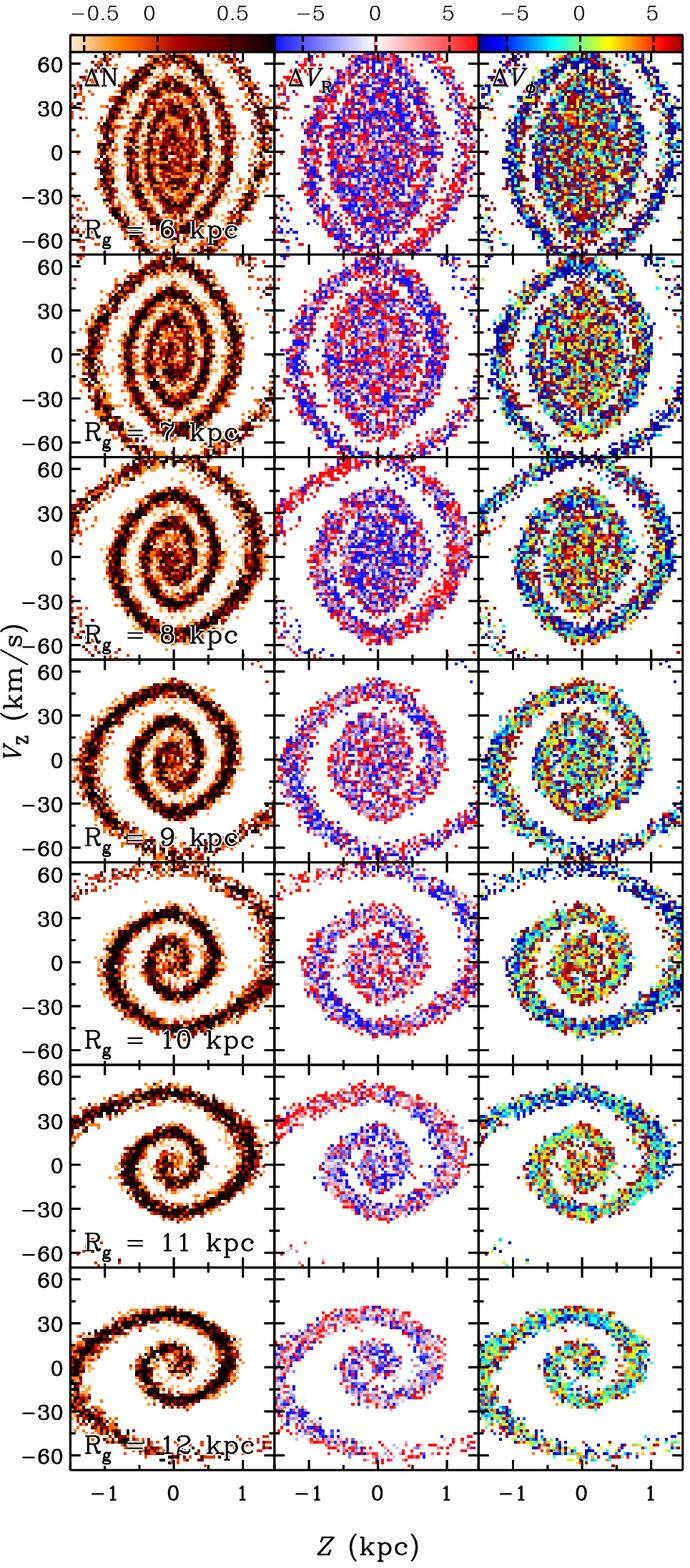}
\caption{The $Z-V_{Z}$ phase space distributions of stars in different radial ranges (left figure) and guiding radius ranges (right figure) of the test particle simulation in the first approach. In the left (right) figure, the first, second, and third columns show the $\Delta N$, $V_{R}$, and $V_{\phi}$ color-coded phase spaces, respectively.}
\epsscale{1.0}
\label{fig:simu}
\end{figure*}

To compare our results with \citet{khanna_etal_19}, we explore the phase space distributions at each $L_{Z}$. We first derive the 6D information for stars in the {\it Gaia} RVS sample using the same solar position and kinematic configuration as \citet{khanna_etal_19}. Then we are able to reproduce Fig.~16 in \citet{khanna_etal_19} by selecting stars in different angular momentum bin with $\Delta L_{Z} = 200$ kpc km/s ($\sim \pm 0.4$ kpc in $R_{g}$). 

Fig.~\ref{fig:snail_lz_200} shows the number density, $\Delta N$, and $L_{Z}$ color-coded phase spaces from top to bottom rows, respectively. Apparently, the snail shell shapes revealed in the top and middle rows ($N$ and $\Delta N$ maps) are quite different from the corresponding $L_{Z}$ color-coded maps in the bottom row. This confirms our previous conclusion that it is the number density map representing the intrinsic phase space structures. The $L_{Z}$ color-coded phase space may not accurately reflect the phase space snail shell structure because it is the number weighted average of the angular momentum of stars. In an imaginary extreme case, where all the stars in each panel of the top row in Fig.~\ref{fig:snail_lz_200} have the same angular momentum, the $L_{Z}$ color-coded $Z-V_{Z}$ phase space will not reveal any feature at all (see Appendix B for more discussion). We can conclude that $R_{g}$-based snail shells can reveal more clear, and intrinsic information on the phase mixing process than the $L_{Z}$ color-coded ones.

\begin{figure*}
\epsscale{1.2}
\plotone{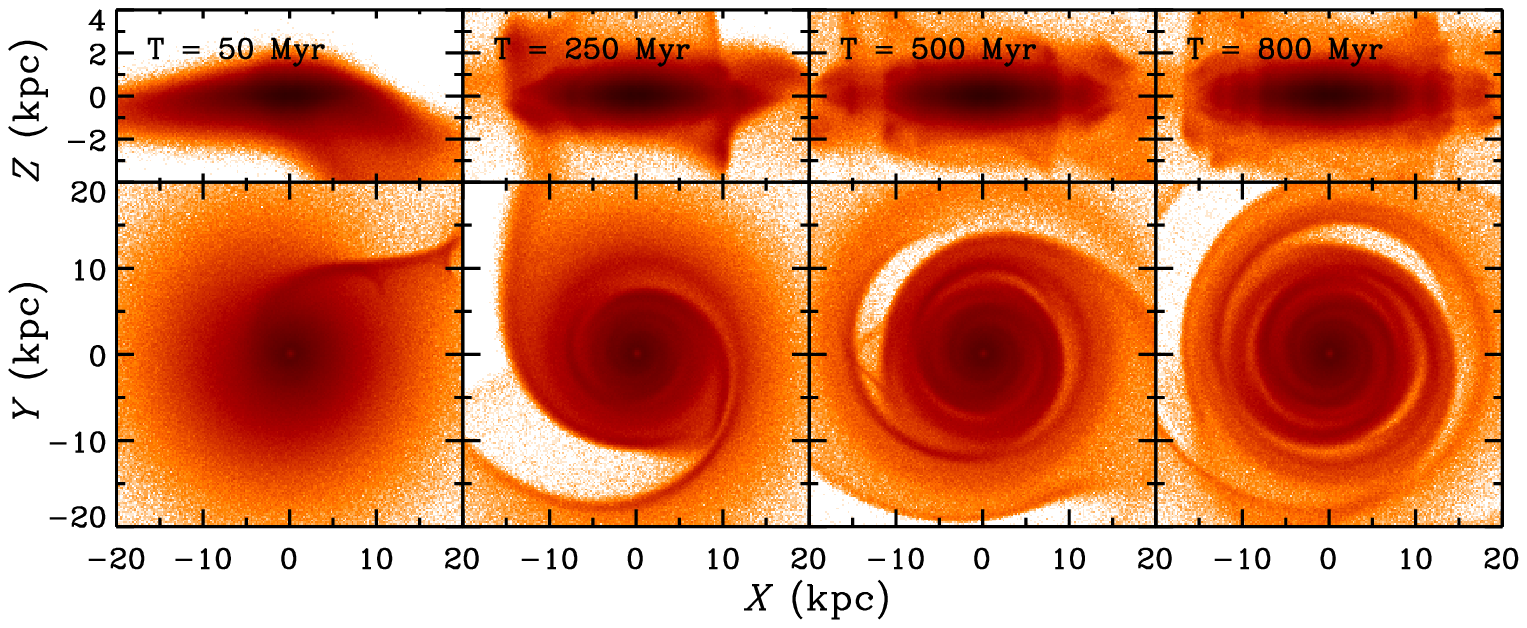}
\caption{The evolution of the test particle simulation with the impulse approximation implemented. The intruder is assumed to hit the disk at 18 kpc along the $X-$axis from the center, with a vertical downward velocity of 300 km/s. From left to right, the four columns show the evolution of the model at $T = 50$, 250, 500, and 800 Myr after the impact. We adopt the snapshot at $T = 500$ Myr in the phase space analysis.}
\epsscale{1.0}
\label{fig:xy_imp}
\end{figure*}

\subsection{Colder and Hotter Orbits Dichotomy}

\citet{li_she_20} found that, for stars near the solar neighborhood, clear phase space snail shell can only be seen in the dynamically colder orbits ($|V_{\phi} - V_{\rm LSR}| < 30$ km/s or $J_{R} < 0.04$), while the hotter orbits have probably phase-wrapped away already. The perturbation should happen at least $\sim$ 500 Myr ago to facilitate the phase mixing of the hotter orbits. Here with a larger sample we also investigate the cold/hot dichotomy across the Galactic disk. Similar to \citet{li_she_20}, at different radius, we use the same criteria to separate the colder and hotter orbits. The phase space distributions for the colder and hotter orbits at different radius are shown in the left and right figures of Fig.~\ref{fig:snail_cold_hot_r}, respectively. Consistent with \citet{li_she_20}, prominent snail shell can only be seen on the colder orbits at different radius. 

We also test to divide stars at each guiding radius into colder and hotter orbits. The $Z-V_{Z}$ phase space distributions are shown in Fig.~\ref{fig:snail_cold_hot_rg} with the colder and hotter orbits shown in the left and right figures, respectively. In fact, we have raised the orbital hotness criteria to $J_{R} > 0.06$, since snail shell features are still visible in the warmer orbit at intermediate $J_{R}$ values ($0.04 < J_{R} < 0.06$). This test again confirms the importance of the guiding radius in revealing the phase space snail shell features, with the radial orbit hotness (ellipticity) playing a less important role.

As shown in Fig.~\ref{fig:snail_cold_hot_rg}, the phase space number density maps of the hotter orbits (right figure) seems to show less vertical excursion than the colder orbits (left figure), inconsistent with the expectation that hotter orbits are more likely to drift away from the disk and reach higher vertical distance. Using $R_{g} = 9$ kpc as an example, we compare the distribution of parameters representing the vertical excursion (i.e., $J_{Z}$, $Z$, and $V_{Z}$) between the colder orbits and hotter orbits\footnote{The colder and hotter orbits in Fig.~\ref{fig:cold_hot_9} are selected with $J_{R} < 0.02$ and $J_{R} > 0.07$, respectively, to enhance the difference in the parameter distributions.}. As shown in the normalized histograms in Fig.~\ref{fig:cold_hot_9}, both the hotter and colder orbits peak at 0, except that the hotter orbits show larger dispersions in $J_{Z}$, $Z$, and $V_{Z}$ than the colder ones, consistent with the expectation. The reason that we found smaller phase space coverage in Fig.~\ref{fig:snail_cold_hot_rg} for the hotter orbits is mainly due to the much lower numbers of stars; at $R_{g} = 9$ kpc, there are 944,231 stars in the colder orbits ($J_{R} < 0.06$), and only 108,095 stars in the hotter ones ($J_{R} > 0.06$). Another possibility of this inconsistency originates from our method to calculate $R_{g}$, where we have ignored the fact that the circular velocity decreases as a function of $Z$ at a given radius. However, this effect should be minor. According to \citet{bov_tre_12}, at the solar radius with $|Z| < 1.5$ kpc, the vertical gradient of the circular velocity can be estimated as $\displaystyle \frac{\partial V_{c}}{\partial Z} \simeq -2.8\ {\rm km\ s^{-1}\ kpc^{-1}} \left(\frac{Z}{100\ \rm pc}\right)$. At $Z = 1$ kpc, the circular velocity is expected to be 14 km/s lower than the circular velocity in the mid-plane. The angular momentum difference is $\sim 110$ kpc km/s, 5\% of the angular momentum at $R_{\odot}$, which should play a minor role in the guiding radius estimation.

\subsection{Vertical Action and Oscillation Frequency Analysis}

Aiming to better understand the vertical phase mixing process, we explore the action space of the sample. Fig.~\ref{fig:omg_jz_n} shows the $\Omega_{Z}-\sqrt{J_{Z}}$ correlations for stars at $R, R_{g} = 8$ kpc. At each radius, the anti-correlation between $\Omega_{Z}$ and $\sqrt{J_{Z}}$ is the main reason behind the phase space snail shell, representing the anharmonic vertical oscillation. Comparing the top and middle panels, the $\Omega_{Z} - \sqrt{J_{Z}}$ distribution at each $R_{g}$ is much tighter than that at different $R$, resulting in a clear snail shell in the number density map of $R_{g}$ selected sample. This also implies that regardless of the current position in the disk, stars with the same angular momentum tend to follow similar $\Omega_{Z}-\sqrt{J_{Z}}$ correlation (and similar snail shell shape in the $Z-V_{Z}$ phase space). 

The top right panel in Fig.~\ref{fig:omg_jz_n} shows the same $\Omega_{Z}-\sqrt{J_{Z}}$ correlation at $R = 8$ kpc but color-coded with $V_{\phi}$. A clear trend can be seen, that $\Omega_{Z}$ decreases for larger $V_{\phi}$ at a given $J_{Z}$, consistent with \citet{bin_sch_18}. On the other hand, the middle right panel shows the $\Omega_{Z}-\sqrt{J_{Z}}$ correlation of stars at $R_{g} = 8$ kpc color-coded with the guiding radius deviation ($\Delta R_{g} = R_{g} - 8$ kpc). The relatively smaller guiding radius corresponds to higher vertical oscillation frequency $\Omega_{Z}$.

To better understand the tight correlation between $\Omega_{Z}$ and $\sqrt{J_{Z}}$ in the $R_{g}$-based sample, we first choose stars in a very narrow guiding radius range, i.e., $R_{g} = 8 \pm 0.01$ kpc, which are further separated into a dynamically colder subsample with $J_{R} < 0.01$ and a dynamically hotter subsample with $J_{R} > 0.06$. The $\Omega_{Z}$ and $\sqrt{J_{Z}}$ distributions of the two subsamples are shown in the bottom left panel of Fig.~\ref{fig:omg_jz_n}. The correlation is very tight for the dynamically colder orbits at very small $J_{R}$, with the dispersion increasing considerably at larger $J_{R}$ for the dynamically hotter orbits. This is expected since the dynamically colder stars are close to circular orbits; these stars with different vertical action move in the vertical gravitational potential well at the same radius, thus resulting in tight correlation between $\Omega_{Z}$ and $J_{Z}$. On the other hand, the dynamically hotter stars have large radial excursion, with the corresponding vertical profile of the gravitational potential varying to result in the large dispersion. Notice that the dispersion is mainly towards lower vertical frequency. This is probably due to the fact that the dynamically hotter stars spend more time around apocenter in the outer disk, with the vertical oscillation frequency reduced in the shallower potential well.

The bottom right panel in Fig.~\ref{fig:omg_jz_n} shows the tight $\Omega_{Z}-\sqrt{J_{Z}}$ correlation for dynamically colder stars at different guiding radius from 6 to 11 kpc, each with 0.02 kpc width annulus and $J_{R} < 0.01$. At a given vertical action value, the vertical oscillation frequency decreases progressively at larger guiding radius. The slope at larger $R_{g}$ is also shallower, consistent with the loosely wound snail shell shown in Fig.~\ref{fig:snail}.

\begin{figure*}
\epsscale{1.}
\plotone{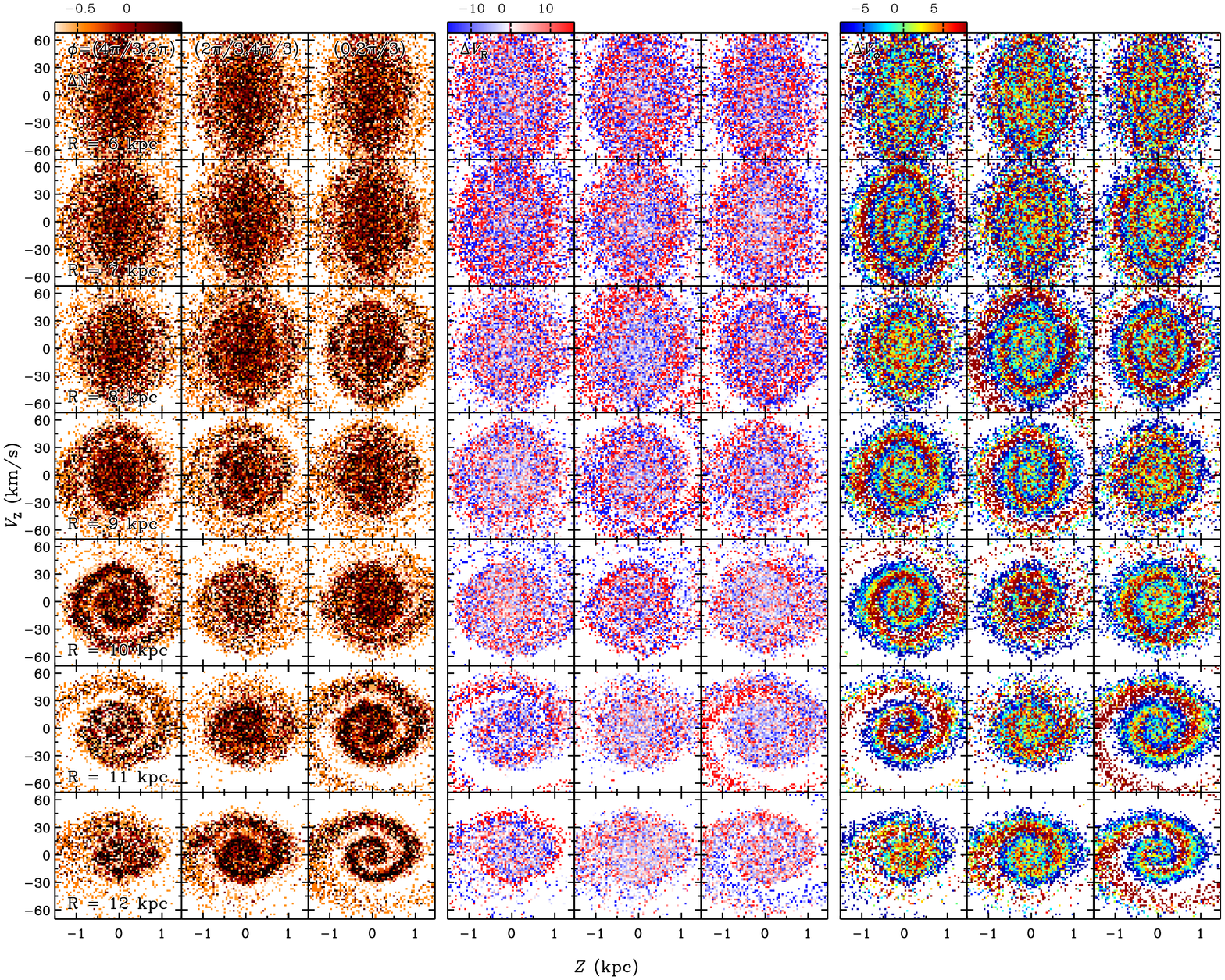}
\caption{The $Z-V_{Z}$ phase space distributions of stars in different radial ranges (from $R = 6$ to 12 kpc) of the test particle simulation in the second approach. The left three columns show the number density maps in the $Z-V_{Z}$ phase space at each radius (different rows) and each azimuthal wedge (different columns).  The middle three columns show the $V_{R}$ color-coded the $Z-V_{Z}$ phase spaces, and the right three columns show the $V_{\phi}$ color-coded phase spaces corresponding to these number density maps.}
\epsscale{1.0}
\label{fig:diff_r_imp}
\end{figure*}

\section{Test Particle Simulations}

To better understand the spatial variation and the importance of the angular momentum of the phase space snail shell across the Galactic disk, we perform test particle simulations in a realistic Milky Way potential, i.e., Model I in  \citet{irrgan_etal_13}. We use AGAMA to construct the stellar distribution function (DF) for the initial condition and to perform the orbit integration of the test particle simulation \citep{vasili_19}. We adopt the {\it QuasiIsothermal} DF for the thin disk \citep{binney_10, bin_mcm_11}, with the radial scale length as 3.7 kpc, and vertical scale height as 0.3 kpc \citep{bin_pif_15, bla_ger_16}.

The vertical perturbation is imposed on the test particles in two different approaches. The first approach is similar to \citet{li_she_20}, where the particles receive vertical downward velocities with the vertical positions barely changed; each particle receives a random vertical velocity perturbation with the median value of $-30$ km/s and dispersion of 5 km/s. For the second approach, we consider a more realistic impulse approximation on the test particle velocities following \citet{bin_sch_18}, to mimic the effect of a fly-by external perturber. Although the test particle simulation is relatively simple, it has the advantage of numerical efficiency and could also reflect the important physical processes of vertical oscillation and phase mixing (see \citealt{bla_tep_20} for a more recent comprehensive numerical modeling with $N$-body simulation).

In the first approach, one million test particles have been sampled from the DF.  Regardless of the different azimuthal angles, the test particles at the same radius receive the same vertical perturbation, and thus the same vertical velocity distributions. In the following analysis, all the particles in each radial (or guiding radial) annulus are used. After 500 Myr evolution, the $Z-V_{Z}$ phase space distributions are shown in Fig.~\ref{fig:simu} for different radial annuli\footnote{To enhance the prominence of the snail shell, we only keep the colder orbits at each radius ($V_{\phi} - V_{\rm circ} < 20$ km/s).} (left figure) and different guiding radial annuli (right figure) similar to Fig.~\ref{fig:snail}. The $R$-based snail shell shapes are similar to $R_{g}$-based snail shell, with the $R_{g}$-based snail shells narrower and well-defined than the $R$-based ones. These results agree with the observational results in Fig.~\ref{fig:snail}. Stars with the same angular momentum tend to follow the same vertical oscillation pattern (although they were located at different radius). On the other hand, in a given radial annulus, stars usually have a wide distribution of the angular momentum. The mixture of stars with different angular momentum (and the related phase space pattern) results in the blurring of the snail shell.

\begin{figure*}
\epsscale{1.}
\plotone{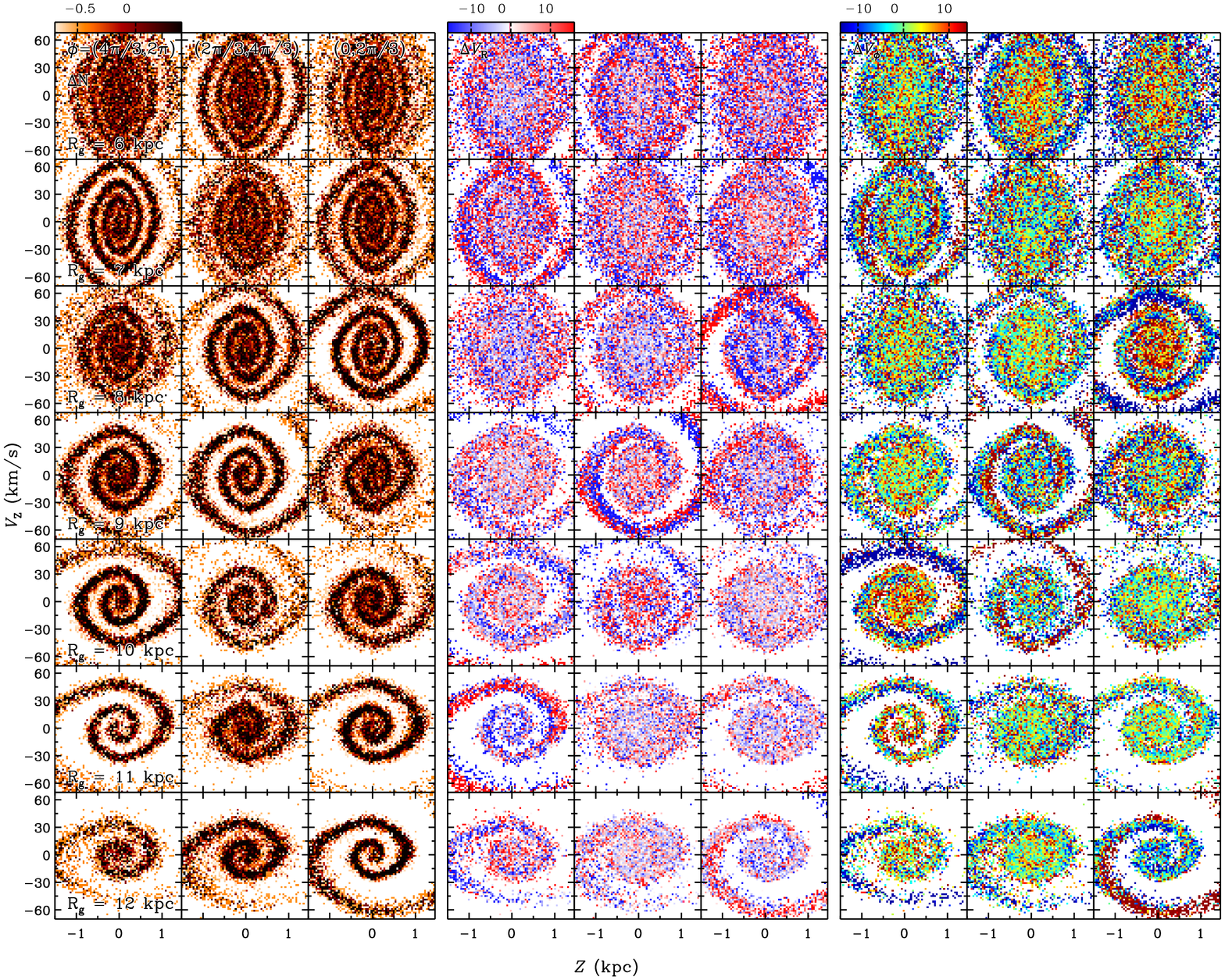}
\caption{The $Z-V_{Z}$ phase space distributions of stars in different guiding radial ranges (from $R_{g} = 6$ to 12 kpc) of the test particle simulation in the second approach. The left three columns show the number density maps in the $Z-V_{Z}$ phase space at each guiding radius (different rows) and each azimuthal wedge (different columns).  The middle three columns show the $V_{R}$ color-coded the $Z-V_{Z}$ phase spaces, and the right three columns show the $V_{\phi}$ color-coded phase spaces corresponding to these number density maps.}
\epsscale{1.0}
\label{fig:diff_rg_imp}
\end{figure*}

\begin{figure}
\epsscale{1.2}
\plotone{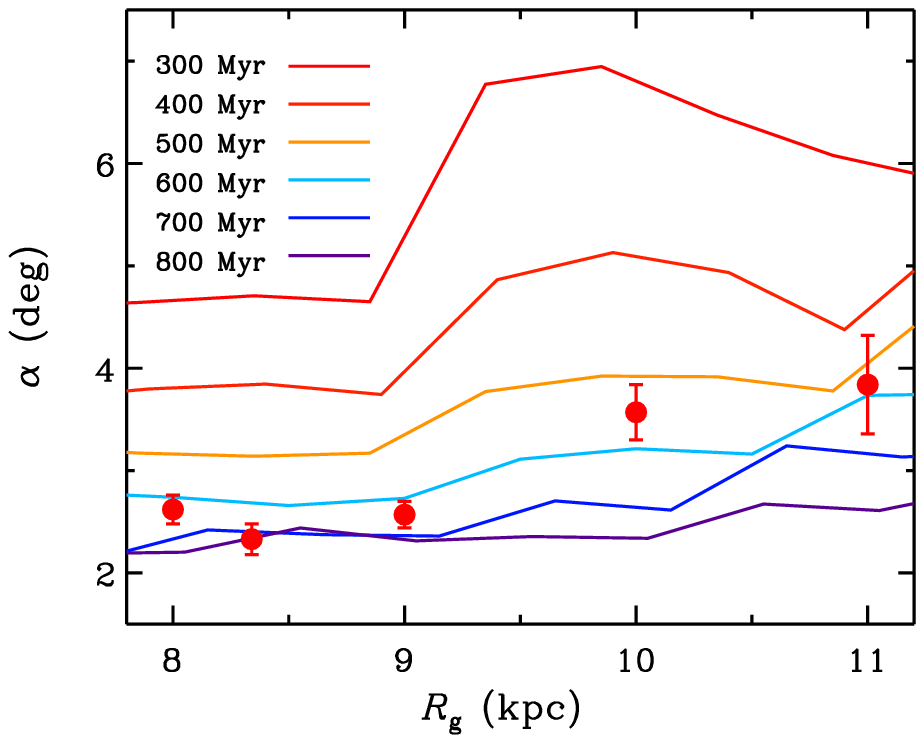}
\caption{Radial profiles of the phase spiral pitch angle of the first test particle simulation with different perturbation times (from 300 Myr to 800 Myr ago). The red points are the observation measurement in Fig.~\ref{fig:pitch_ang_prof}. The pitch angle profiles with the perturbations from 500 to 700 Myr ago agree well with the observation.}
\label{fig:pitch_comp}
\epsscale{1.}
\end{figure}

In the left figure of Fig.~\ref{fig:simu}, the third column shows the $V_{\phi}$ color-coded phase space. Different color shows different snail shell patterns. For particles at a given radius, the azimuthal velocity is proportional to the angular momentum. Since stars with different angular momentum have different snail shell  pattern, when color-coded with $V_{\phi}$, different snail shells naturally emerge in the phase space. Comparing the results with different $R_{g}$ ranges in the right figure, there is no systematic variation of the snail shell shape with azimuthal velocity, since the median $V_{\phi}$ at different part of the phase space along the snail shell is close to 0. In Fig.~\ref{fig:simu} for both the $R$- and $R_{g}$-based phase spaces, no variation of the snail shell pattern with $V_{R}$ can be seen. The median radial velocity in the phase space is close to 0. This is mainly due to the initially imposed symmetric distribution of $V_{R}$ in our test particle simulation.

For the second approach, we have sampled three million test particles based on the DF to account for the azimuthal variation of the phase space snail shell. The perturbation to the vertical velocities of the particles are estimated following the impulse approximation in \citet{bin_sch_18}. The impact parameter of the external perturber is 18 kpc along the $X-$axis from the Galactic center at a speed of $\sim$ 300 km/s. The mass of the perturber is $2\times10^{10} M_{\odot}$ and $T = 66$ Myr as the passage timescale. To work in the non-inertial frame with the Galactic center stationary, the gravitational pull from the intruder to a point mass at the Galactic center is subtracted from the acceleration of the intruder on the other test particles. The in-plane velocity change $\delta v_{\parallel}$ is computed by multiplying the characteristic timescale of the passage with the acceleration. A downward component $\delta v_{\perp}$ is added as $\delta v_{\perp} = \alpha \delta v_{\parallel} [1 - \beta \sin{(\psi - \psi_{\rm intruder})}]$ with $\alpha = 1.5$ and $\beta = 0.5$ \citep[see Eqt. 3 in][]{bin_sch_18}\footnote{We have increased $\alpha$ from 0.4 to 1.5 to enhance the vertical perturbation effect.}. The vertical perturbation is composed of two components: a constant term associated with the standard dynamical friction and a term proportional to $\sin{(\psi - \psi_{\rm intruder})}$. The second term arises from the consideration that stars initially moving towards the intruder  ($\psi - \psi_{\rm intruder} < 0$) receive a net downward kick, while those stars initially moving away from the intruder ($\psi - \psi_{\rm intruder} > 0$) receive a net upward kick. We define $\psi_{\rm intruder} = 0^\degree$ along the $X-$ axis, and the azimuthal angle increases counterclockwise.

After the impact, the time evolution of the test particles is shown in Fig.~\ref{fig:xy_imp}. As shown in the first column, 50 Myr after the perturbation, the disk becomes lopsided that is stretched towards the intruder with a prominent single arm. The amplitude of the vertical perturbation of each particle depends on the azimuthal angle, with larger amplitude for particles closer to the intruder. This is the main difference compared to the previous test particle simulation where the imposed vertical perturbation is azimuthally symmetric. In the second column, 250 Myr after the impact, the initial perturbation gradually evolves to form the $m = 2$ mode due to the differential rotation, with the vertical oscillation in different parts of the disk. After 500 Myr evolution (third column), the $m = 2$ mode is still present, with continuous phase mixing in the vertical direction of the disk. In fact, this is the snapshot we adopt in the following analysis, since the phase space snail shell is most prominent and roughly similar to the observations. At later times, as shown in the fourth column (800 Myr), the face-on image still exhibit prominent $m = 2$ mode feature, with the roughly symmetric distribution of the particles with respect to the disk mid-plane in the edge-on view of the model.

Recently, \citet{bla_tep_20} used a high resolution N-body model to simulate the impact of an external intruder on the main disk. They found good agreement with the model in \citet{bin_sch_18} indicating that the impulse approximation is valid to describe such process. After the impact, \citet{bla_tep_20} noticed a strong, $m = 1$ bending mode across the disk is set up, which gradually wraps up due to the differential rotation of the disk to result in a superposition of two distinct bisymmetric ($m = 2$) modes, namely a spiral pattern and a bending wave. Their self-consistent N-body simulation results are quite similar to our simple test particle simulation under the impulse approximation. 

Since the vertical perturbation is not azimuthally symmetric, different phase space snail shells are expected to emerge at different azimuthal angles. The disk is separated into three azimuthal wedges with $120^\degree$ each, in order to trace the azimuthal variation of the phase space snail shell with sufficient number of particles. The $Z-V_{Z}$ phase space distributions at different radius and azimuthal wedges are shown in Fig.~\ref{fig:diff_r_imp}. At each radius, as expected, the shape of the phase space snail shell is different in different azimuthal wedge. For example, at $R = 8$ and 9 kpc, the snail shell is barely seen in the azimuthal wedge with $\phi = (4\pi/3, 2\pi)$ (first column), while in the other azimuthal wedges at $(2\pi/3, 4\pi/3)$ (second column) and $(0, 2\pi/3)$ (third column), the snail shells are clear. In the $V_{R}$ (middle three columns) and $V_{\phi}$ (right three columns) color-coded phase spaces, evidence of snail shell shape changing at different velocities can be seen, roughly consistent with observations.

The phase space distributions at different guiding radius are shown in Fig.~\ref{fig:diff_rg_imp}. With azimuthal variations still present, the $R_{g}$-based number density phase space (left three columns) show prominent snail shells compared to the $R$-based phase spaces in Fig.~\ref{fig:diff_r_imp}. In the $V_{R}$ and $V_{\phi}$ color-coded phase spaces, there is no systematic variation of the snail shell shape with the velocities. \citet{bla_tep_20} commented that they could not find the phase space snail shell feature as a function of the angular momentum. Still, their model is one of the best available in the literature that fits the global Milky Way properties well.

Based on the first test particle simulation, the pitch angles of the phase space snail shells in different guiding radii ($8 \sim 11$ kpc) at different times ($300 \sim 800$ Myr) are measured following the method in Section 4.1. By comparing to observational pitch angles, this could help to date the perturbation event; the pitch angle decreases monotonically as the phase space snail shell winds up with time. The radial profiles of the measured pitch angle at different times are shown in Fig.~\ref{fig:pitch_comp}. The pitch angle decreases from $\sim 6^\degree$ at 300 Myr to $\sim 2.2^\degree$ at 800 Myr, consistent with our expectation for tighter snail shells at earlier perturbation events. The observational result seems to agree well with the pitch angle profiles from 500 to 700 Myr, with the best agreement at T = 600 Myr. This constrain of the perturbation event is also consistent with previous results using other different methods, such as the consecutive turns of the snail shell \citep{antoja_etal_18}, the cold/hot orbit dichotomy of the snail shell \citep{li_she_20}, and vertical potential reconstruction with oscillation frequency estimation \citep{li_wid_21}.

\section{summary}

We utilize $\sim$ 11 million stars from both the {\it Gaia} RVS sample and the {\it LAMOST} DR6 sample to study the vertical phase mixing across the Galactic disk. We confirm the existence of the snail shells in the $Z-V_{Z}$ phase space in the Galactic disk from 6 to 12 kpc, and quantitatively measure the shape of the snail shells, with the corresponding pitch angles also derived. We find that the guiding radius (angular momentum) is fundamental to reveal the vertical phase mixing signals. Compared to the $R$-based phase spaces, the clarity of the snail shell in the $R_{g}$-based phase space is increased; more wraps of the snail shell can be seen; phase space is less affected by the lack of stars close to the disk mid-plane due to extinction; snail shell signal is also amplified in a greater radial range.

In the epicycle theory, a star revolves around the guiding radius that is on circular orbits in the disk plane. When perturbed in the vertical direction, the amplitudes and the frequency of the stellar vertical oscillation depend on the shape of the vertical potential at different radius. Along its orbit, as the star moves inside $R_{g}$, the vertical oscillation frequency generally increases with lower $Z_{\rm max}$ and higher $V_{Z\rm max}$, vice versa for the star moving outside $R_{g}$. Averaging along the orbits, the $Z-V_{Z}$ phase space trajectories of the stars with the same angular momentum would generally show a coherent pattern with small deviations. Therefore, grouping stars into different $R_{g}$ helps to enhance such phase space signal with more prominent snail shell.

Additional evidence comes from the cold/hot dichotomy discovered in \citet{li_she_20} in the solar neighborhood. For the $R$-based phase space, the cold/hot dichotomy still holds, with prominent snail shell only in the colder orbits. However, in each $R_{g}$ range, snail shell can still be seen in warmer orbits. The snail shell eventually disappears for very hot orbits with $J_{R} > 0.06$ (compared to the original criteria $J_{R} > 0.04$ for the hotter orbits in \citealt{li_she_20}). Compared to the colder orbits, the warmer orbits ($0.04 < J_{R} < 0.06$) show larger radial excursion. Within each $R_{g}$ range, they still show similar snail shell feature as the colder orbits, indicating that the vertical perturbation is probably global and external origin. If the perturbation happens in the inner disk, the orbits with the same $L_{Z}$ but different radial excursion will likely encounter the outward propagating wave at different times, to show differences in the initial phase angle of the snail shell, which is likely to blur the phase space signals. Therefore, to study the vertical phase mixing process, it is recommended to use $R_{g}$ (with additional constraints on $J_{R}$) to better reveal the intrinsic shape of the snail shell in the Galactic disk. In another ongoing work, we are trying to model the snail shell shapes at different radius to constrain the shape of the Galactic potential.

The difference between the phase space distributions in the Galactocentric radial range and the guiding radius range can also be understood in the $\Omega_{Z}-\sqrt{J_{Z}}$ plane. At each radius, the distribution of stars in the $\Omega_{Z}-\sqrt{J_{Z}}$ plane shows anti-correlation with a broad distribution, while stars in the corresponding $R_{g}$ range show a much tighter correlation, corresponding to more prominent snail shells in the phase space.

Test particle simulations are also performed in two approaches to understand this phenomenon. In the first approach, a simple azimuthally symmetric vertical perturbation is imposed on all the test particles. Snail shell features can be seen across the disk, which are more prominent when the particles are grouped into different guiding radial ranges. In the second approach, the perturbation with the impulse approximation is applied on all the test particles. Again, the phase space snail shell is more prominent in different guiding radial ranges, consistent with the observations. By comparing the pitch angle profiles between the observation and the simulation, the perturbation event is constrained to happen at $\sim$ 500 to 700 Myr ago, consistent with other studies using different methods.

I thank the referee for helpful comments and suggestions that improved the quality of the paper. I also thank Juntai Shen for insightful suggestions that help improve the clarity of the paper, and Chao Liu for helpful discussions. The research presented here is supported by the ``111'' Project of the Ministry of Education under grant No. B20019, by the National Natural Science Foundation of China under grant No. 11621303, and by the MOE Key Lab for Particle Physics, Astrophysics and Cosmology. This work made use of the Gravity Supercomputer at the Department of Astronomy, Shanghai Jiao Tong University, and the facilities of the Center for High Performance Computing at Shanghai Astronomical Observatory.

Guoshoujing Telescope (the Large Sky Area Multi-Object Fiber Spectroscopic Telescope {\it LAMOST}) is a National Major Scientific Project built by the Chinese Academy of Sciences. This work has made use of data from the European Space Agency (ESA) mission {\it Gaia} (http://www.cosmos.esa.int/gaia), processed by the {\it Gaia} Data Processing and Analysis Consortium (DPAC, http://www.cosmos.esa.int/web/gaia/dpac/consortium). Funding for the DPAC has been provided by national institutions, in particular the institutions participating in the {\it Gaia} Multilateral Agreement.


\begin{appendix}

\section{A. $V_{\phi}$ Color-Coded Phase Space Properties}

\begin{figure}
\epsscale{1.}
\plotone{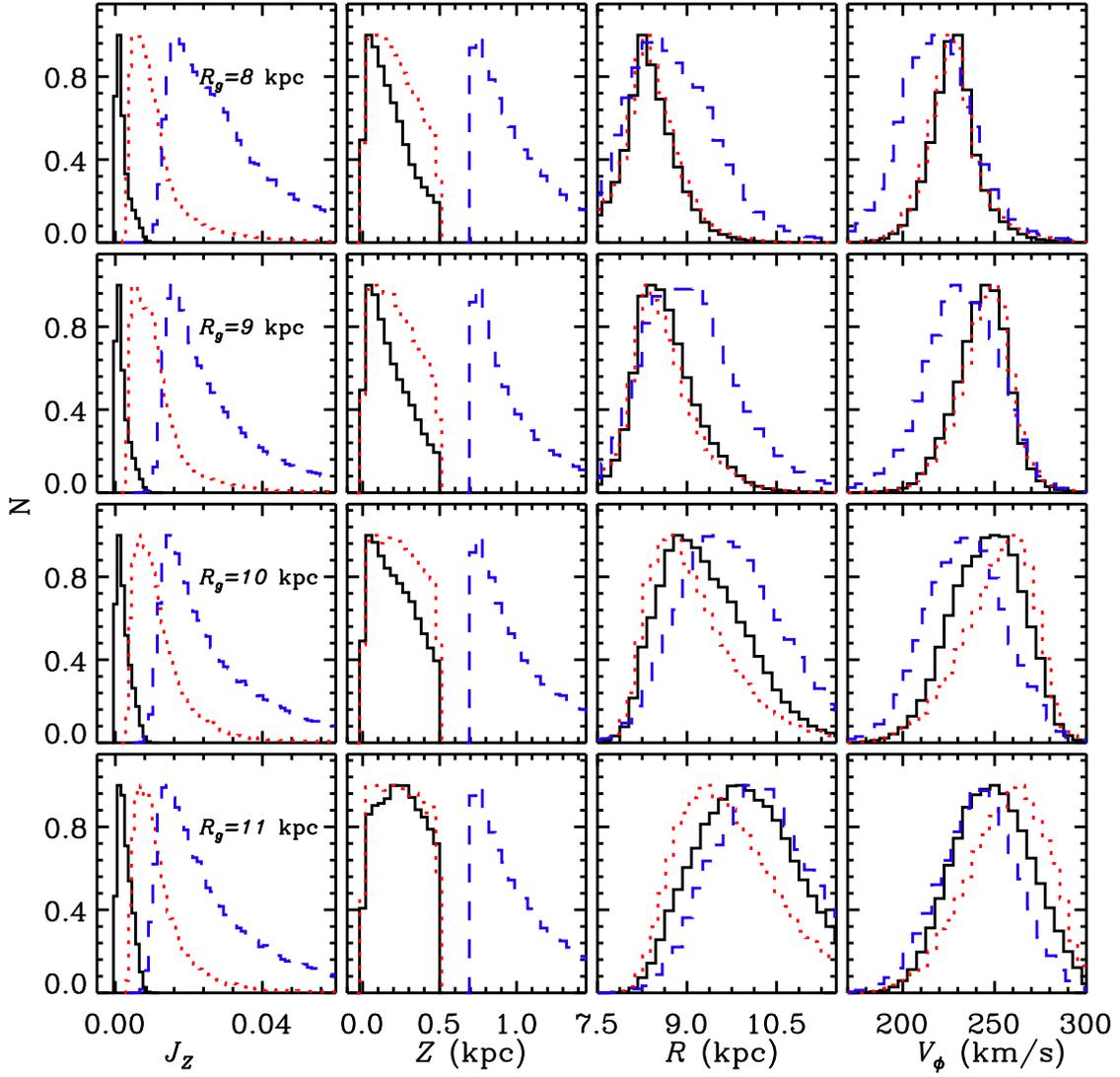}
\caption{Distributions of vertical action $J_{Z}$, vertical height $Z$, radius $R$ and azimuthal velocity $V_{\phi}$ of stars with the guiding radius $R_{g} = 8, 9, 10$, and 11 kpc as shown from the top to bottom rows. At each guiding radius, three subsamples are selected with the black sold line, red dotted line, and the blue dashed lines representing subsample 1 ($|Z| \leq 0.5$ kpc and $|V_{Z}| \leq 15$ km/s), subsample 2 ($|Z| \leq 0.5$ kpc and $|V_{Z}| \geq 20$ km/s), and subsample 3 ($|Z| \geq 0.7$ kpc), respectively.}
\label{fig:hist}
\end{figure}

As shown in the right figure in Fig.~\ref{fig:snail}, in the $V_{\phi}$ color-coded phase space at $R_{g} = 8, 9, 10, 11,$ and (possibly) 12 kpc, the average azimuthal velocity seems smaller at larger $|Z|$ and higher at larger $|V_{Z}|$. This feature can be understood from the statistical distribution of stellar orbital properties. Focusing on $R_{g} = 8, 9, 10,$ and 11 kpc, we select three subsamples at each guiding radius bin, with subsample 1 at $|Z| \leq 0.5$ kpc and $|V_{Z}| \leq 15$ km/s, subsample 2 at $|Z| \leq 0.5$ kpc and $|V_{Z}| \geq 20$ km/s, and subsample 3 at $|Z| > 0.7$ kpc. Fig.~\ref{fig:hist} shows the distributions of $J_{Z}$, $Z$, $R$ and $V_{\phi}$ at each guiding radius, with the black solid line, the red dotted line and the blue dashed line representing the subsamples 1, 2, and 3, respectively. As expected from the location in the $Z-V_{Z}$ diagram, and also shown in the first column, the vertical action is smallest in subsample 1, and largest in subsample 3. The vertical height is largest in subsample 3 (second column), corresponding to larger radial range shown in the third column; stars with larger vertical hights tend to be located outwards at larger radius where the disk potential well is shallower. At each guiding radius, these three subsamples have similar angular momentum. Therefore, the larger radii naturally result in smaller azimuthal velocities in subsample 3 (fourth column). 

For subsample 2, although the vertical action is higher than that of subsample 1 as shown in the first column, they show similar vertical height distributions (second column), with subsample 1 slightly narrower. The much larger vertical velocities of subsample 2 hint for smaller radius for this subsample; stars are more likely to acquire larger vertical velocities when traveling at smaller radii with deeper potential well. This picture is consistent with the third column. Therefore, the azimuthal velocity of subsample 2 is relatively higher than that of subsample 1.

Besides the $V_{\phi}$ pattern of the phase space at each $R_{g}$, careful inspection of the left columns in Fig.~\ref{fig:snail} for the snail shell at different radial annuli indicates differences between the snail shell shapes of the $\Delta N$ map (first column) and $V_{\phi}$ color-coded phase spaces (third column), especially for the annulus at 10 kpc. This difference is likely caused by the snail shell shape variation at different $V_{\phi}$. According to \citet{li_she_20}, from the mathematical point of view, the $V_{\phi}$ color-coded phase space can be considered as the number weighted average of the azimuthal velocity of stars in each radius. With $V_{\phi}$ color-coding, problems with the incomplete sampling close to the mid-plane can be alleviated. However, the $V_{\phi}$ color-coded phase space may not truly reflect the real phase space snail shell shape. \citet{li_she_20} has shown in their Fig.~14 that by combining stars following different snail shell shapes at slightly different $V_{\phi}$, the phase space color-coded with $V_{\phi}$ would look quite different from the $\Delta N$ map. As shown in the second and third columns of the right figure for different guiding radius ranges, there is even no well recognized snail shells at all for the $V_{R}$ and $V_{\phi}$ color-coded phase spaces.

\section{B. $Z-V_{Z}$ Phase Space at Different $R_{g}$}

\begin{figure}
\epsscale{1.2}
\plotone{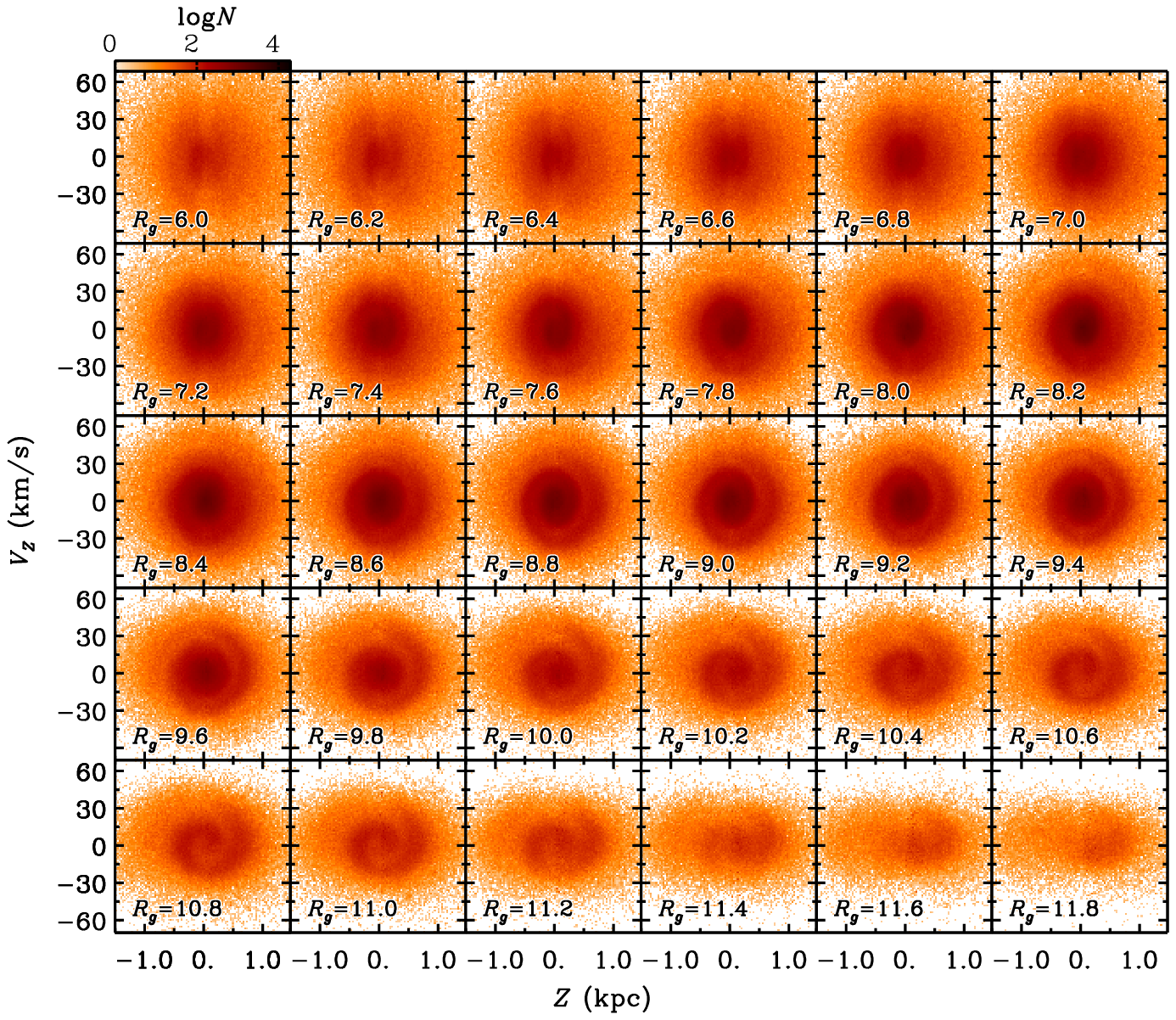}
\caption{The number density map $N$ of the $Z-V_{Z}$ phase space distributions for stars with different guiding radius $R_{g}$.}
\epsscale{1.0}
\label{fig:zvz_lz_n}
\end{figure}

\begin{figure}
\epsscale{1.2}
\plotone{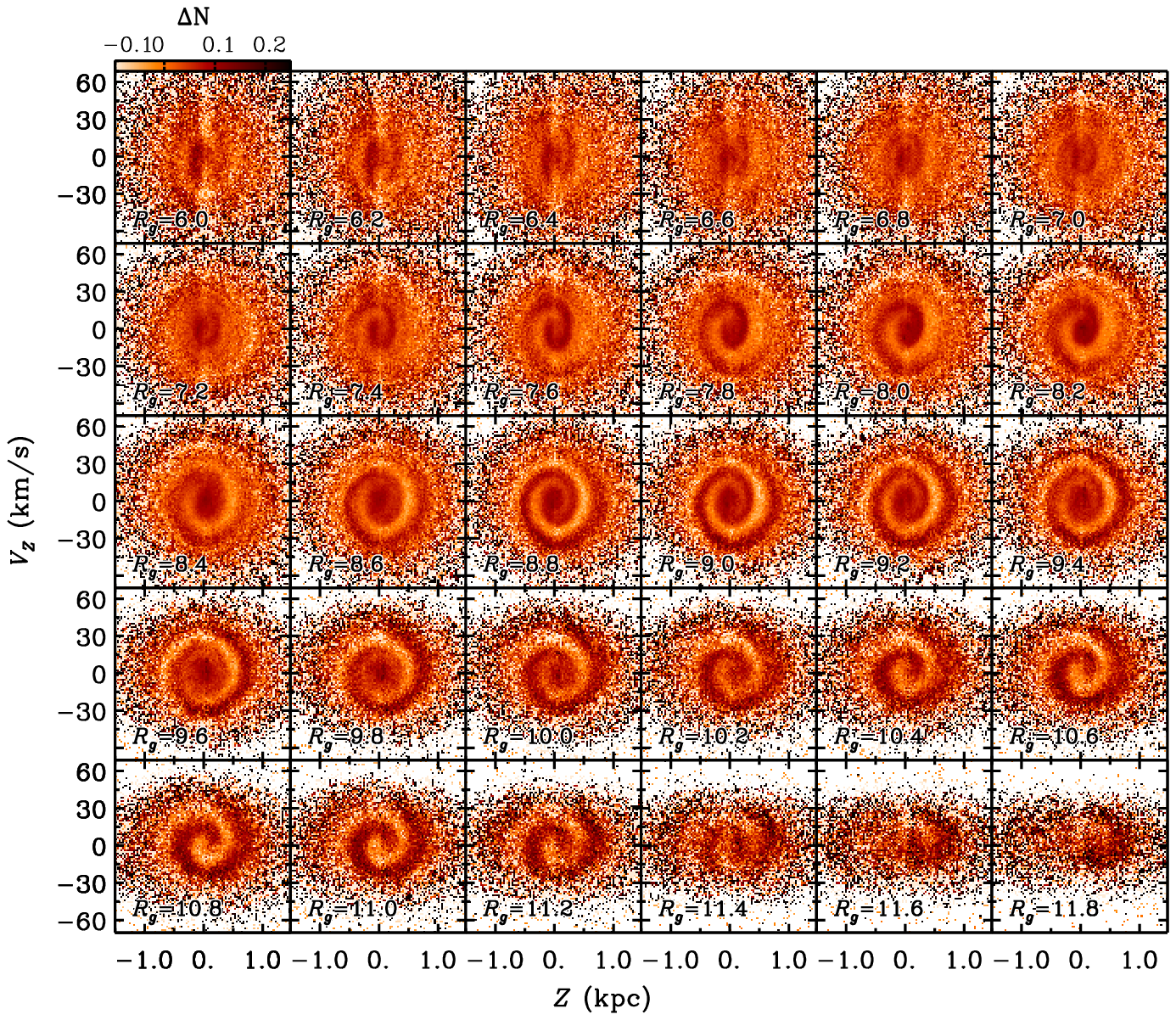}
\caption{The number density contrast map $\Delta N$ of the $Z-V_{Z}$ phase space distributions for stars with different guiding radius $R_{g}$.}
\epsscale{1.0}
\label{fig:zvz_lz_dn}
\end{figure}

\begin{figure}
\epsscale{1.2}
\plotone{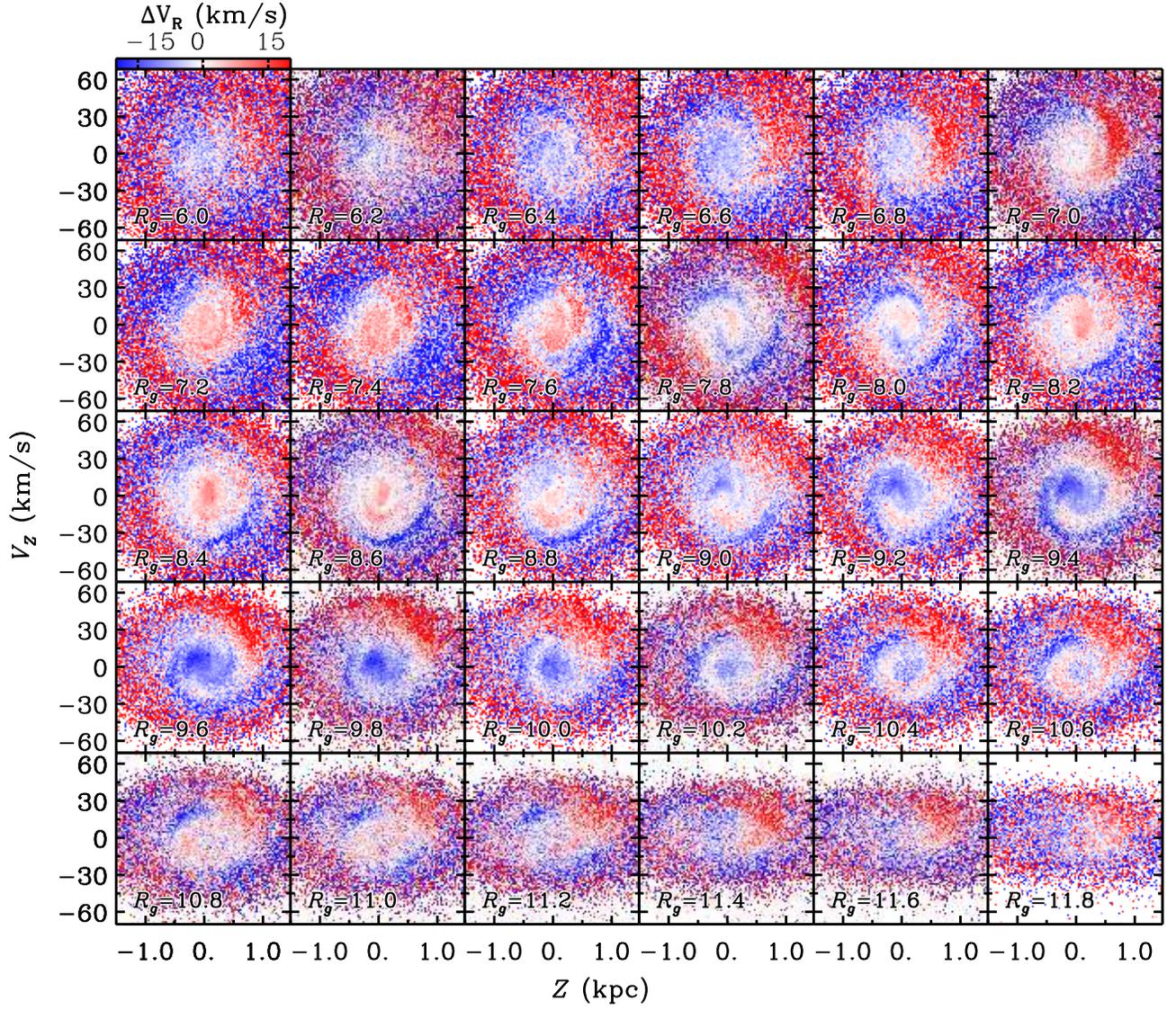}
\caption{The radial velocity ($V_{R}$) color-coded $Z-V_{Z}$ phase space distributions for stars with different guiding radius $R_{g}$.}
\epsscale{1.0}
\label{fig:zvz_lz_vr}
\end{figure}

\begin{figure}
\epsscale{1.2}
\plotone{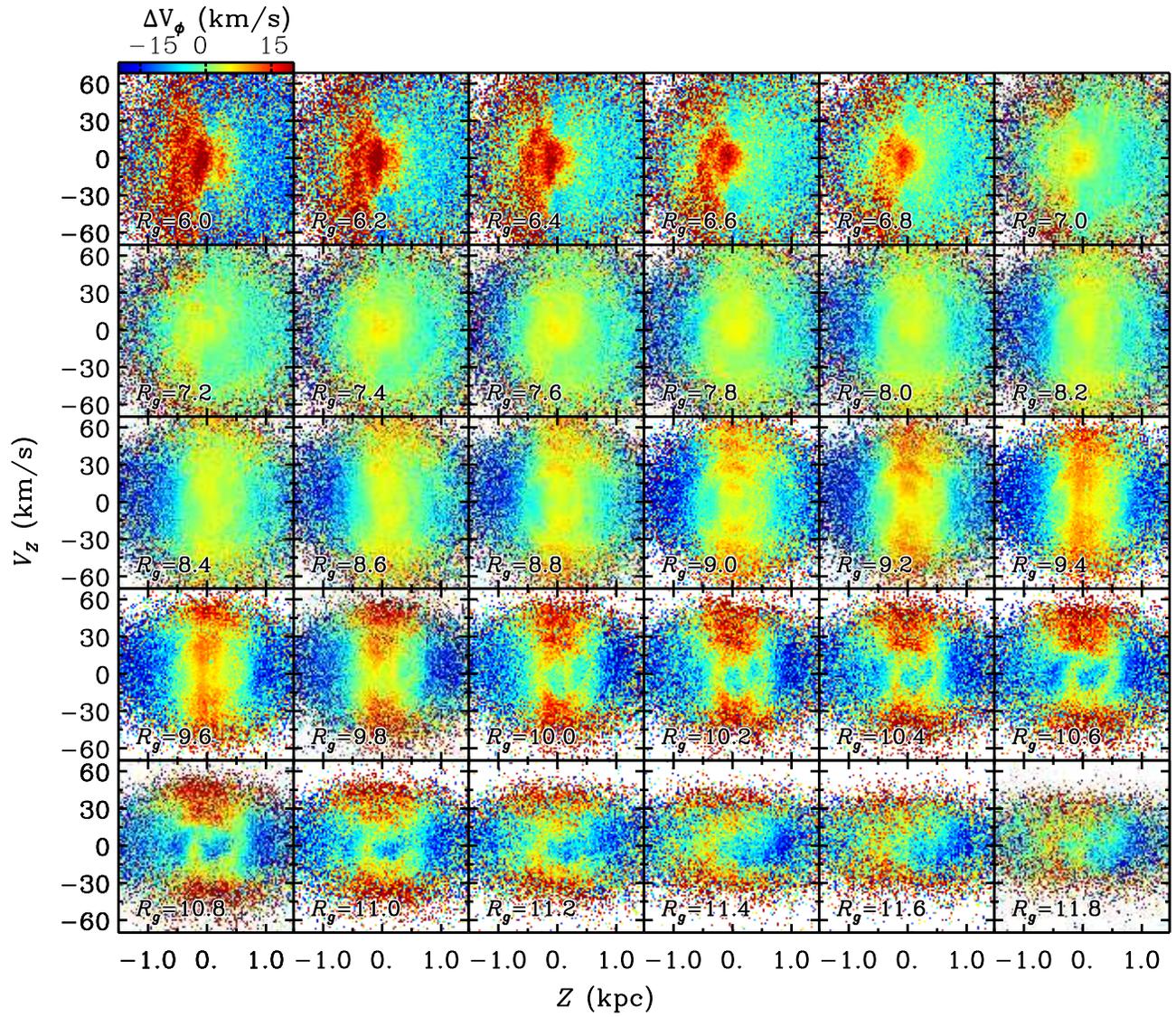}
\caption{The azimuthal velocity ($V_{\phi}$) color-coded $Z-V_{Z}$ phase space for stars with different guiding radius $R_{g}$.}
\epsscale{1.0}
\label{fig:zvz_lz_vphi}
\end{figure}

\begin{figure}
\epsscale{1.2}
\plotone{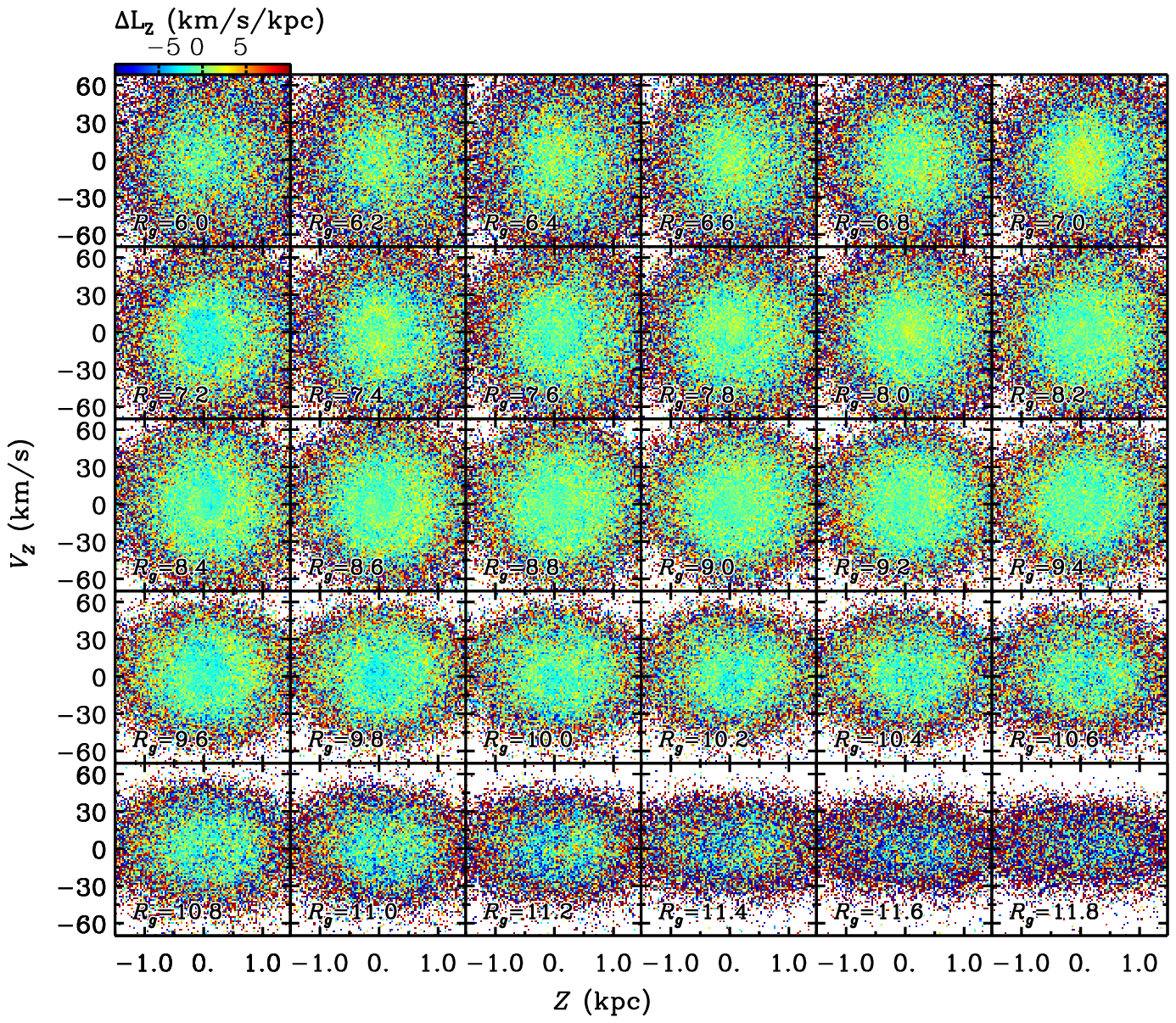}
\caption{The angular momentum $L_{Z}$ color-coded $Z-V_{Z}$ phase space for stars with different guiding radius $R_{g}$.}
\epsscale{1.0}
\label{fig:zvz_lz_lz}
\end{figure}

In order to better visualize the gradual transition of the phase space snail shell across the Galactic plane, we separate the sample into 30 guiding radius bins from 6 to 12 kpc with the bin width of 0.2 kpc. The number density map and the contrast maps are shown in Figs.~\ref{fig:zvz_lz_n} and~\ref{fig:zvz_lz_dn}, respectively. As the guiding radius increases, the snail shell becomes less wound and more extended along the $Z$ direction. The gradual transition is very clear. The $V_{R}$, $V_{\phi}$, and $L_{Z}$ color-coded phase spaces are shown in Figs.~\ref{fig:zvz_lz_vr},~\ref{fig:zvz_lz_vphi}, and~\ref{fig:zvz_lz_lz}, respectively. No clear pattern can be seen, consistent with our previous results.

\section{C. The Gap in the Phase Space around $Z=0$}

\begin{figure}
\epsscale{1.}
\plotone{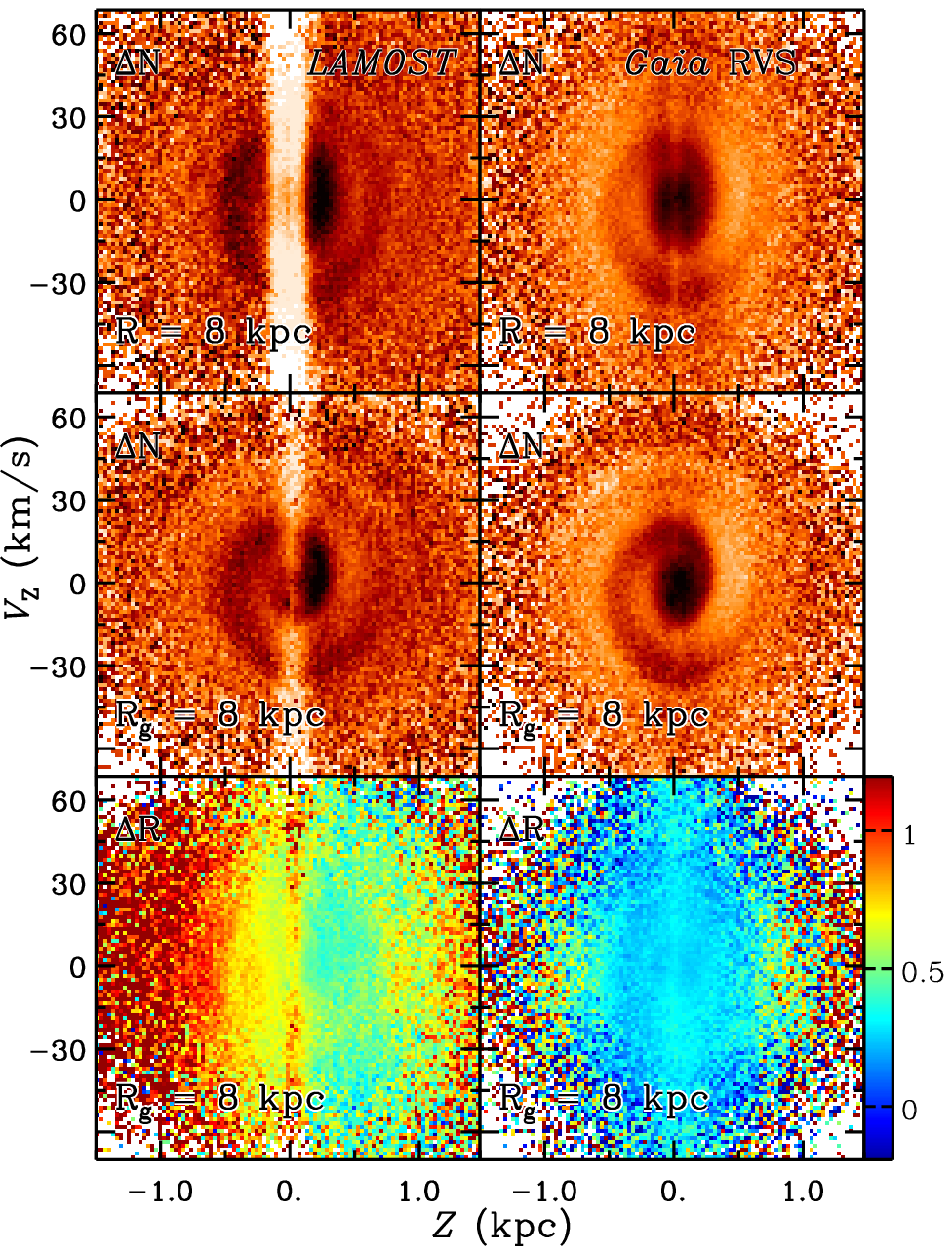}
\caption{The number density contrast map $\Delta N$ of the $Z-V_{Z}$ phase space distributions for stars in the {\it LAMOST} and {\it Gaia} RVS samples at 8 kpc. The left and right panels show the {\it LAMOST} and {\it Gaia} RVS results, respectively. The top and middle rows show the phase space map at the radius $R$ and guiding radius $R_{g}$, respectively. The bottom row is the phase space at $R_{g}$ and color-coded with $\Delta R = R - 8$ kpc.}
\epsscale{1.0}
\label{fig:mid_gap_8}
\end{figure}

\begin{figure}
\epsscale{1.}
\plotone{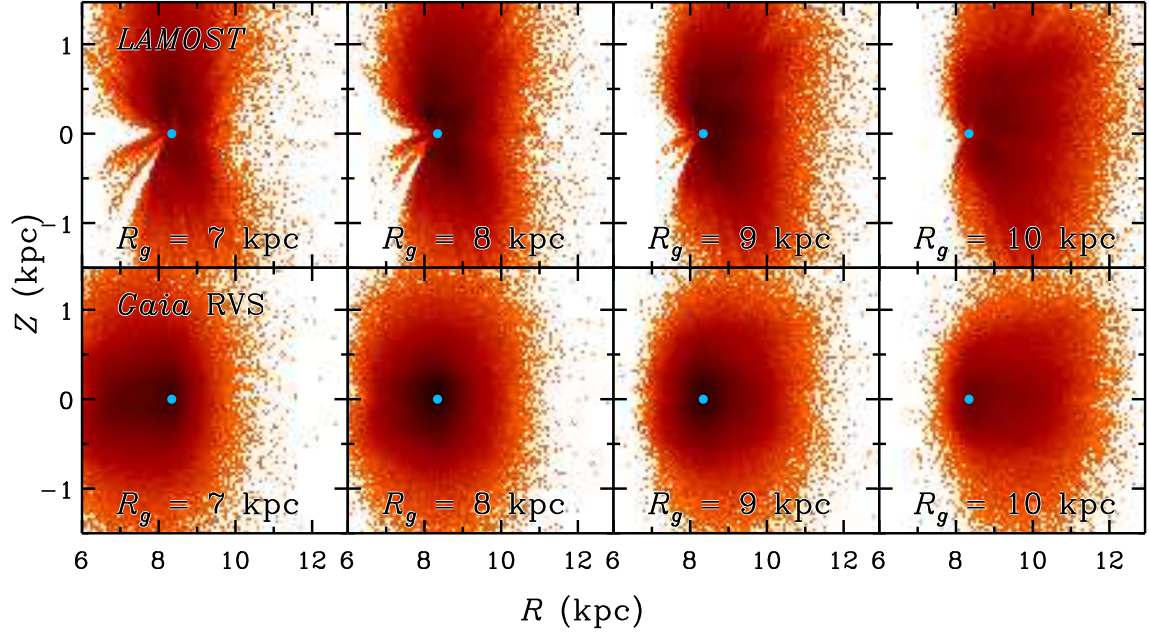}
\caption{The $R-Z$ distributions for stars at different guiding radii ($R_{g}$) in the {\it LAMOST} (top row) and {\it Gaia} samples (bottom row). The cyan dot marks the position of the Sun. }
\epsscale{1.0}
\label{fig:rz_rg}
\end{figure}

In Fig.~\ref{fig:snail}, a clear gap close to $Z = 0$ in the $R$-based phase space number density map can be seen, especially at radius smaller than 8 kpc. This feature is mainly caused by the stronger dust attenuation in the disk mid-plane. However, this gap becomes much weaker or even disappears in the corresponding phase space at $R_{g}$. Here we perform detailed comparison between the {\it LAMOST} and {\it Gaia} RVS sample at each $R$ and $R_{g}$ to better understand this difference. Fig.~\ref{fig:mid_gap_8} shows the comparison at 8 kpc. The gap is very clear in the {\it LAMOST} sample at $R = 8$ kpc (top row), and much weaker at $R_{g} = 8$ kpc (middle row). For the {\it Gaia} RVS sample, the gap is weaker, but the improvement in $R_{g}$ is apparent. The bottom panel shows the same $R_{g}$ phase space as the middle row but color-coded with $\Delta R = R - 8$ kpc. It seems that at $R_{g} = 8$ kpc, most stars come from $R = 8\sim9$ kpc. In this case, stars in the disk are reshuffled with a large fraction of stars near the solar radius (where the observed stellar number density is highest close to the mid-plane) redistributed to other $R_{g}$ values. Fig.~\ref{fig:rz_rg} shows the $R-Z$ distribution of stars with guiding radii $R_{g} = 7$ to 10 kpc in the {\it LAMOST} (top row) and {\it Gaia} samples (bottom row). The contribution from stars near the solar radius is significant at each guiding radius. These stars are also well sampled across the vertical direction, especially close to the mid-plane around $Z = 0$. In fact, stars at each guiding radius origin from a large radial range, with a high contribution especially from 8 to 9 kpc. 

\end{appendix}

\end{document}